\documentclass[10pt]{article}
\usepackage{titlesec}
\usepackage[utf8]{inputenc}
\usepackage[T1]{fontenc}
\usepackage{amsmath}
\usepackage{graphicx}
\usepackage{braket}
\usepackage{amsfonts}
\usepackage{amssymb}
\usepackage{geometry}
\usepackage{xfrac} 
\usepackage{physics} 
\usepackage{enumerate}
\usepackage{setspace}
\usepackage{multicol} 
\usepackage{xcolor} 
\usepackage{float}
\usepackage[nottoc,numbib]{tocbibind}
\usepackage{multirow} 
\usepackage{lscape}
\usepackage{diagbox}
\usepackage{hyperref}
\usepackage{subfig}
\usepackage{yfonts}
\usepackage{mathrsfs}
\usepackage{ulem}
\usepackage{tabularx} 
\usepackage{bm}
\usepackage{mathtools}
\usepackage{amsbsy}

\usepackage[scr=bickham]{mathalfa}

\usepackage[mathscr]{euscript} 

 \usepackage{verbatim} 

\newcommand{\paren}[1]{\left( #1 \right)}
\newcommand{\loga}[1]{ \log{\left(#1\right)}}
\newcommand{\logacorche}[1]{ \log{\left[#1\right]}}
\newcommand{\corche}[1]{\left[#1\right]}
\newcommand{\llave}[1]{ \left \{ #1\right \}}
\newcommand{\mean}[1]{\left<#1\right>}

\renewcommand{\l}{l}
\newcommand{\dif}{\text{d}}

\newcommand{\dt}{\text{d}t}
\newcommand{\dr}{\text{d}r}

\newcommand{\expo}[1]{\mathrm{e}^{#1}}
\newcommand{\pe}{\alpha}
\newcommand{\Ye}{Y}
\newcommand{\Ot}{\Omega_{n,m}^{\lambda,\omega_c}}

\newcommand{\Otnl}{\Omega_{n,m}^{\lambda,\omega_c}}
\newcommand{\Otol}{\Omega_{0,m}^{\lambda,\omega_c}}
\newcommand{\Otoo}{\Omega_{0,0}^{\lambda,\omega_c}}

\newcommand{\normNla}{\mathcal{N}_{n,m}^{\lambda,\omega_c}}

\newcommand{\G}[1]{\Gamma\left(#1\right)}

\newgeometry{bottom=2.5cm, top=2.5 cm, left=2.5cm, right=2.5cm}

\title{The Fock-Darwin-Darboux system: eigenstates, information entropies and dispersion-like measures}

\begin{document}

\maketitle

\begin{center}

{\sc Ignacio Baena-Jimenez$^{1}$, Angel Ballesteros$^{1}$ and Ivan Gutierrez-Sagredo$^{2}$}

\medskip
{$^1$Departamento de F\'isica, Universidad de Burgos, 
09001 Burgos, Spain}

{$^2$Departamento de Matem\'aticas y Computaci\'on, Universidad de Burgos, 
09001 Burgos, Spain}

\medskip
 
e-mail: {\href{mailto:ibaena@ubu.es}{ibaena@ubu.es}, \href{mailto:angelb@ubu.es}{angelb@ubu.es}, \href{mailto:igsagredo@ubu.es}{igsagredo@ubu.es}}

\end{center}

\begin{abstract}

The Fock-Darwin (FD) quantum system describes the motion on the plane of a charged particle under the action of an isotropic oscillator potential together with a perpendicular constant magnetic field. When the isotropic oscillator is suppressed, the FD system leads to the Landau Hamiltonian with infinitely degenerate Landau levels. The Fock-Darwin-Darboux (FDD) system is the generalisation of the FD system to a particle moving on the Darboux III space, which is a conformally flat surface with non-constant negative curvature. We present a systematic study of some information-theoretic entropy and dispersion-like measures for these quantum systems. Since both systems are exactly solvable, analytical expressions for Shannon, Rényi and Tsallis entropies, among others, can be obtained. We show that for the FD system, its information-theoretic measures are formally the same as the ones for the harmonic oscillator, provided a modified effective frequency depending on the magnetic field is introduced. In the FDD case, the nonlinear nature of the underlying manifold precludes the existence of a simple closed form for the wave-function on momentum space, which is numerically analysed. We compare the numerical behaviour of the different entropy measures and we analyse the interplay arising in the FDD system between the curvature parameter and the magnetic field. In particular, it is shown that the Landau system on the Darboux III space has no infinitely degenerate Landau levels.

\end{abstract}

\bigskip

\noindent
KEYWORDS: Shannon entropy; Rényi entropy; Tsallis entropy; quantum information; nonlinear quantum oscillator; Darboux III oscillator; Fock-Darwin system; Landau levels; curvature.

\renewcommand{\listfigurename}{LISTA DE FIGURAS}
\renewcommand{\listtablename}{Lista de Tablas}
\renewcommand{\figurename}{Fig. }
\renewcommand{\tablename}{Table} 

\tableofcontents{}

\section{Introduction}

Quantum systems coupled with magnetic fields have been found to play a central role in both fundamental physics and emerging technologies. For instance, in quantum information and computation, magnetic interactions are routinely employed to manipulate and control qubits (see \cite{lu2023quantum} for a recent review). Magnetic fields are also essential for exploring exotic states of matter, with the quantum Hall effect standing as a paradigmatic example in condensed matter physics \cite{chang2023colloquium,li20213d}. Finally, coupling to magnetic fields underpins powerful applications in quantum sensing, ranging from ultra-sensitive magnetometry capable of detecting extremely weak fields \cite{gottscholl2021spin,tang2021high} to biosensing approaches with direct relevance to biomedical diagnostics \cite{bao2023quantum,zadeh2022magnetic}. 

Two particularly interesting examples including magnetic fields, both of them exactly solvable, are the two-dimensional systems known as the Fock-Darwin and the Darboux III oscillators. The Fock-Darwin system \cite{fock1928bemerkung,darwin1931diamagnetism} describes a two-dimensional harmonic oscillator coupled to a constant perpendicular magnetic field $B$, and its Hamiltonian is given by
\begin{align}
    \mathcal{H}_{FD}=\frac{1}{2m} \paren{\hat{p}_x + \frac{e B}{2 c} \hat{y}}^2+\frac{1}{2m} \paren{\hat{p}_y - \frac{e B}{2 c} \hat{x}}^2+\frac{k}{2 } \paren{\hat{x}^2
    +\hat{y}^2}.
\end{align}
Beyond its mathematical interest due to its symmetries and associated integrability properties (see~\cite{drigho2017superintegrability} for a review, including the classical dynamics of the FD system), its $k\to 0$ limit is just the Landau system describing the motion of a charged particle within a constant perpendicular magnetic field and giving rise to Landau levels with infinite degeneracy. As a consequence, the FD system has outstanding physical applications. For instance, it provides the standard description of charged particles confined in two-dimensional semiconductor quantum dots under an external magnetic field \cite{babinski2006emission,makarovsky2008fock,henriques2009magnetoconfined}, where it has been widely used to analyse spectral properties, shell structure and transport effects \cite{hawrylak1993single,kouwenhoven2001few,johnson2000enhanced}. Extensions of the model have also been considered for anisotropic quantum dots and for systems including additional interactions such as Zeeman, Rashba or external electric-field terms \cite{madhav1994electronic,avetisyan2012strong}. Moreover, the Fock-Darwin model continues to appear naturally in recent works, where it emerges as part of the theoretical framework or as a limiting case in the analysis of nanoscale systems
\cite{chen2025manifesting,leon2020coherent,pena2020quasistatic}.
Also, closely related Fock--Darwin-type Hamiltonians appear in the study of coherent states and in paraxial optics, where analogous structures describe Hermite--Gaussian and Laguerre--Gaussian modes (see~\cite{drigho2017superintegrability}).

On the other hand, the two-dimensional version of the Darboux III quantum Hamiltonian is given by
\begin{align}
    \mathcal{H}_{D}=\frac{1}{2\paren{1+\lambda (\hat{x}^2+\hat{y}^2)}} \paren{\hat{p}_x^2 + \hat{p}_y^2+\frac{k}{2} \paren{\hat{x}^2
    +\hat{y}^2} }\, ,
    \label{DIII}
\end{align}
with $\lambda \in \mathbb R^+$. In arbitrary dimensions, the Darboux III oscillator \cite{ballesteros2008maximally,ballesteros2011new,ballesteros2011quantum} can be viewed as 
an exactly solvable superintegrable  $\lambda$-deformation of the isotropic harmonic oscillator.
It is an $N$-dimensional maximally superintegrable system representing an oscillator defined on a radially symmetric space with non-constant negative curvature, corresponding to the $N$-dimensional generalisation of the so-called Darboux surface of type III \cite{darboux1915leccons}, a surface with nonconstant negative curvature defined by the conformally flat metric
\begin{equation}
\dif s^2 = (1+ \lambda (x^2 + y^2)) (\dif x^2 + \dif y^2)\, ,
\end{equation}
whose $\lambda\to 0$ limit leads to flat space. It is worth stressing that in General Relativity the Darboux III metric was shown in~\cite{Ballesteros:2008xu} to arise as the spatial part of one of the so-called Bertrand  spacetimes~\cite{Perlick_1992}, which are spherically symmetric and static Lorentzian spacetimes in which any bounded trajectory is periodic. Alternatively, the two-dimensional  Darboux III oscillator~\eqref{DIII} can  be also interpreted as a position-dependent mass (PDM) system on the plane, with mass function given by $\mu(x,y)=1+\lambda (x^2+ y^2) $ \cite{ballesteros2008maximally,ballesteros2011new,ballesteros2011quantum}. As it is well-known, PDM systems are relevant in different physical contexts, such as semiconductor physics, quantum-confined systems, gravitational or cosmological models, among others (see~\cite{el2020generalized,sarker2017position,zhao2020influence,morris2015new,ballesteros2017hamiltonians} and references therein). We stress that although both interpretations, as a system on a curved space and as a PDM system on the plane, are completely equivalent, we will hereafter mainly use the first one.

This paper is based on the fact that the FD system and the Darboux III oscillator can be coupled, since a magnetic field can be added to the latter by preserving their exact solvability provided that its Hamiltonian is defined as \cite{ballesteros2024dunkl}
\begin{align}
    \mathcal{H}_{FDD}=\frac{1}{2\paren{1+\lambda (\hat{x}^2+\hat{y}^2)}} \paren{\paren{\hat{p}_x + \frac{e B}{2 c} \hat{y}}^2+ \paren{\hat{p}_y - \frac{e B}{2 c} \hat{x}}^2+\frac{k}{2} \paren{\hat{x}^2
    +\hat{y}^2} }\, .
    \label{FDD}
\end{align}
In this way, the system~\eqref{FDD} can be viewed as an integrable $\lambda$-deformation of the Fock-Darwin oscillator, or more precisely, as the Fock-Darwin quantum oscillator on the Darboux III surface. Hereafter we will call this Hamiltonian the Fock-Darwin-Darboux system (FDD), whose  properties will be analysed in detail throughout this paper. We stress that this nonlinear quantum model contains three independent parameters: the curvature (or PDM) parameter $\lambda$, the magnetic field $B$ and the oscillator strength $k$, whose rich interplay can be thoroughly studied due to the exact solvability of the Hamiltonian. Moreover, the $k\to 0$ limit of the FDD system leads to the Landau-Darboux Hamiltonian, whose degeneracy properties will be shown to be deeply influenced by the curvature of the underlying space.

The study of entropic and related information-theoretic measures in quantum systems has received considerable attention, including analyses based on Rényi-type quantities, deformed models and phase-space localisation properties \cite{bayindir2021self,akhshani2012image}. As is well known, entropic measures provide valuable insights into the localisation, spreading and complexity of the corresponding wavefunctions \cite{lopez2015statistical,hall1999universal,bialynicki2006formulation,zozor2007classes,puertas2017heisenberg}. 
We recall that Shannon entropy \cite{shannon1948mathematical} is one of the most widely used measures of information and has become a cornerstone of numerous applications in information theory \cite{thomas2006elements,gray2011entropy}. Rényi \cite{renyi1961measures} and Tsallis \cite{tsallis1988possible} entropies have been introduced as generalisations of the Shannon entropies depending on a parameter $\pe$. The properties and applications of these one-parameter generalisations of the Shannon entropy have also been thoroughly investigated (see, for instance, \cite{aczel1975measures,jizba2004world,leonenko2008class,sen2011statistical,jizba2015role,olendski2020renyi} and references therein).
In the specific context of magnetic interactions, several works have explored the interplay between entropy-like descriptors and electric and magnetic-field effects \cite{olendski2015comparative,liang2016fractal,bitane2010magnetic,turkyilmazoglu2018analytical}. Likewise, entropic and dispersion measures for oscillator-type quantum systems in the presence of magnetic fields have been investigated in different settings \cite{xavier2011renyi,hab2022magnetic,wakamatsu2020physics,sjue2015radial}. 

For all these reasons, the aim of this paper is to present a complete study of the entropy-based and dispersion measures for the FD and FDD Hamiltonians, and particular emphasis will be devoted to Rényi and Tsallis entropies. All these aspects are unexplored yet and provide a useful insight into the quantum behaviour of both systems. For the FDD model the role of the nonlinearity parameter $\lambda$ and of the magnetic field $B$ will be studied in detail, as well as the interplay between them. Indeed, the $B\to 0$ limit of the FDD system leads to the Darboux III oscillator, while the $\lambda\to 0$ limit of the FDD Hamiltonian leads to the FD system, and all the results presented in the paper behave smoothly under these two limits.
We recall that for the N-dimensional Darboux III oscillator (without magnetic field) its Shannon entropy was derived in~\cite{ballesteros2023shannon} and the Rényi and Tsallis entropies of the corresponding one-dimensional model have been recently presented in~\cite{baena2025renyi}.

We outline the structure and main results of the paper. Section~\ref{sec: Fock-Darwin quantum system} reviews the Fock-Darwin system and shows that the corresponding entropic and dispersion measures follow directly from the well-known results for the harmonic oscillator. These quantities provide the limiting case of the two-dimensional Darboux III oscillator in a magnetic field as $\lambda\to 0$. In Section \ref{sec: Magnetic Darboux III oscillator}, we introduce and describe the geometric features of the Darboux III surface, which turns out to be a smooth deformation of the hyperbolic space around the origin and becomes asymptotically flat at large distances. Then we define and analyse the generalisation of the Fock-Darwin system on the Darboux III surface, the FDD system, whose spectrum, wavefunctions and accidental degeneracies are explicitly presented and studied in detail. In particular, eigenstates and eigenvalues for the FDD system will be presented in terms of the curvature parameter $\lambda$, and it will be shown that the Landau-Darboux system for any $\lambda\neq 0$ can have at most one infinitely degenerate Landau level.
The entropic and dispersion measures of the FDD system are derived in Section~\ref{sec: resultados FDD}. In particular, we obtain closed-form expressions in position space for the entropic moment~\eqref{eq: entropic moment magnetic darboux} and for the Rényi \eqref{eq: renyi magnetic darboux} and Tsallis \eqref{eq: tsallis magnetic darboux} entropies. The momentum-space entropic measures, as well as the entropy-based uncertainty principle, are studied numerically (see Figure~(\ref{grid: entropias incertidumbre})). We also derive analytical expressions for the dispersion measures in position space, Eqs.~\eqref{eq: rk darboux} and~\eqref{eq: r2 darboux}, and in momentum space, Eqs.~\eqref{eq: mean p2 Darboux III magnetico} and~\eqref{eq: mean p2 l magnetico}. The dispersion-based uncertainty principle is likewise analysed analytically; see Figure~\ref{grid:dispersion measures}. Section~\ref{sec: Magnetic field regimes of special interest} examines the influence of the magnetic field in several regimes of interest. We show that the magnetic field and the curvature parameter $\lambda$ may counteract each other, and we determine the Larmor frequency for which the quadratic position expectation value, Eq.~\eqref{eq: omega cut en r}, coincides with that of the harmonic oscillator. We also consider the case of an inverted magnetic-field direction and derive the corresponding relation for the Larmor frequency, Eq.~\eqref{eq: relacion invertir campo}. Section~\ref{Conclusions magnetico} summarises the main results of the work and points to some open problems directly related to it.

\section{The Fock-Darwin (FD) quantum system} \label{sec: Fock-Darwin quantum system}

In this section, we first review the exact solutions and properties of the FD system, which are directly connected with the ones for the standard isotropic harmonic oscillator. Afterwards we present their information entropies and dispersion measures and, where applicable, we analyse the influence of the magnetic field in the corresponding uncertainty principles.

\subsection{Hamiltonian and densities of the Fock-Darwin system}

The Fock–Darwin quantum system \cite{fock1928bemerkung,darwin1931diamagnetism} is an exactly solvable model describing a charged particle confined to a two-dimensional plane, subject to a harmonic potential and coupled to a magnetic field perpendicular to that plane. Although the system is always solvable, it is not in general superintegrable. As shown in \cite{drigho2017superintegrability}, superintegrability occurs only for special values of the parameters, namely when the ratio of the two relevant frequencies is rational; the isotropic harmonic-oscillator and Landau limits arise as particular superintegrable cases. We can describe the system explicitly in terms of Cartesian coordinates $\{x,y,z\}$ on $\mathbb R^3$, where the particle moves in the $xy$-plane and the magnetic field points in the $z$-direction.
Using the symmetric gauge for the vector potential, we write
\begin{align} \label{eq: potencial vector}
    \boldsymbol{A}=\paren{-\frac{B}{2}y,+\frac{B}{2}x,0}, \hspace{1cm} \boldsymbol{B}=\paren{0,0,B}.
\end{align}
With the previous conventions, the Fock-Darwin Hamiltonian is given by
\begin{align}
    \mathcal{H}=\frac{1}{2m} \paren{\hat{p}_x + \frac{e B}{2 c} \hat{y}}^2+\frac{1}{2m} \paren{\hat{p}_y - \frac{e B}{2 c} \hat{x}}^2+\frac{k}{2 } \paren{\hat{x}^2
    +\hat{y}^2},
\end{align}
where $\hat{p}_x$ and $\hat{p}_y$ are momentum operators
\begin{align}
    \hat{p}_x=-i \hbar \pdv{}{x}, \hspace{1cm} \hat{p}_y=-i \hbar \pdv{}{y},
\end{align}
and the frequency of the oscillator is given by $\omega^2 = \frac{k}{m}$. From now on we will set $m = 1$, and therefore this Hamiltonian can be rewritten as
\begin{align}
\mathcal{H}=\frac{1}{2}\paren{\left(\hat{p}_x^2+\hat{p}_y^2\right)+\omega_c^2\left(\hat{x}^2+\hat{y}^2\right)-2 \omega_c \hat{L}_z}+\frac{1}{2} \omega^2\left(\hat{x}^2+\hat{y}^2\right),
\end{align}
where $\omega_c=\frac{e B}{2 c}$ is the Larmor frequency and $\hat{L}_z=\hat{x} \hat{p}_y-\hat{y} \hat{p}_x$ is the angular momentum in the $z$ direction. Thus, introducing a modulation total frequency defined as
\begin{align} \label{eq: omega t}
    \omega_t=\sqrt{\omega_c^2 + \omega^2},
\end{align}
the Hamiltonian becomes 
\begin{align}
\mathcal{H}=\frac{1}{2} \paren{\left(\hat{p}_x^2+\hat{p}_y^2\right)+\omega_t^2\left(\hat{x}^2+\hat{y}^2\right)-2 \omega_c \hat{L}_z} .
\end{align}
This form of the Hamiltonian yields the same wave functions as those of the isotropic harmonic oscillator, but with an effective angular frequency $\omega_t$. The energy levels are given by
\begin{align}
    E_{n,m}^{\omega_c}=\hbar \omega_t (2n+\abs{m}+1)-\hbar m \omega_c, \hspace{1cm} n=0,1,2,\dots \hspace{0.5 cm} m=0,\pm1,\pm 2,\dots 
\end{align}

A brief comment on the degeneracy of the spectrum and its physical significance is in order. To this aim we introduce the dimensionless energy $\epsilon_{n,m}^\nu$, as discussed in \cite{drigho2017superintegrability}, which makes the dependence on the magnetic field particularly transparent, and is given by
\begin{align}
    \epsilon_{n,m}^\nu=\frac{E_{n,m}^{\omega_c}}{\hbar \omega_t}=(2n+\abs{m}+1)- \nu m,
\end{align}
where the Fock-Darwin dimensionless frequency is 
\begin{align} \label{eq: nu}
    \nu=\frac{\omega_c}{\omega_t}=\frac{\omega_c}{\sqrt{\omega^2+\omega_c^2}}.
\end{align}
For $\nu=0$ ($\omega_c=0$) we recover the harmonic oscillator system, whereas $\nu=1$ ($\omega = 0$) gives rise to the Landau system \cite{landau1965collected}, that is, a charged particle moving on a plane with a uniform magnetic field perpendicular to it. For generic values of $\nu$, the levels are non-degenerate. Degeneracies appear only for special values of the parameter $\nu$, namely when two different pairs of quantum numbers $(n_1,m_1)$ and $(n_2,m_2)$ give the same value of the dimensionless energy (see Figure \ref{fig: dimensionless energy}). For 
\begin{align} \label{eq: c}
    c = 2 n + \abs{m}+1,
\end{align}
the condition for degeneracy translates to 
\begin{align} \label{eq: nu FD}
    \epsilon^\nu_{n_1,m_1}=\epsilon^\nu_{n_2,m_2},
    \qquad
    \nu=\frac{c_1-c_2}{m_1-m_2}=\frac{q_1}{q_2}, 
\end{align}
with $q_1 \in \mathbb Z$, $q_2 \in \mathbb Z^*$ such that $|q_1| \leq |q_2|$ and $q_1 q_2 \geq 0$. Thus, degeneracies occur at isolated rational values of $\nu$, for which the spectrum exhibits finite degeneracy. The limiting case $\nu=0$, since it corresponds to the isotropic harmonic oscillator, is also finitely degenerate. For the Landau limit $\nu=1$, the energy becomes independent of one quantum number, giving rise to infinitely degenerate Landau levels.

In polar coordinates $(r,\varphi)$ defined by $x=r \cos \varphi, y = r \sin \varphi$, the wave functions of the FD system read
\begin{align}
\Psi_{n,m}^{\omega_c}(r,\varphi)&=R_{n,m}(r) Y_{m}(\varphi), 
\end{align}
where the radial and angular parts are given, respectively, by 
\begin{align}
R_{n,m}^{\omega_c}(r)&=\mathcal{N}_{n,m}^{\omega_c} \, r^\l \, e^{-\frac{\beta_t^2}{2} r^2} \, L_n^{\l} \left( \beta_t^2 r^2 \right), \\
Y_{m}(\varphi)&=\frac{1}{\sqrt{2\pi}} \expo{i m \varphi }, 
\end{align}
with $l=\abs{m}$, $\beta_t=\sqrt{\frac{ \omega_t}{\hbar}}$ and where the normalization constant of the radial part is
\begin{align} \label{eq: normalizacion laguerre radial}
\mathcal{N}_{n,m}^{\omega_c} =
\sqrt{\frac{2 n! \ \beta_t^{2l+2}}{\Gamma \paren{ n + l + 1}}}.
\end{align}
For $\hbar=1$, the probability density becomes 
\begin{align} \label{eq: rho FD}
\rho_{n,m}^{\omega_c}(r)&=\frac{ n! \ \omega_t^{l+1}}{\pi \Gamma \paren{ n + l + 1}} \, r^{2\l} \, e^{-\omega_t r^2} \, \paren{L_n^{\l}\paren{ \omega_t r^2}}^2.
\end{align}
We can get an analytical expression for the wave function of the Fock-Darwin oscillator on momentum space by Fourier transforming the wave function above, and we obtain
\begin{align}
\tilde{\Psi}_{n,m}^{\omega_c}(p,k_\varphi)&=\mathcal{\tilde{N}}_{n,m}^{\omega_c} \, p^\l \, e^{-\frac{p^2}{2\omega_t} } \, L_n^{\l} \left(  \frac{p^2}{\omega_t} \right) \frac{1}{\sqrt{2\pi}}\expo{i m k_\varphi } \, ,
\end{align}
where $(p,k_\varphi)$ are the conjugate coordinates of $(r,\varphi)$.
The normalization constants of the radial part in position and momentum spaces are related by 
\begin{align} \label{eq: norm fock darwin momento}
    \mathcal{\tilde{N}}^{\omega_c}_{n,m}=\frac{\mathcal{N}^{\omega_c}_{n,m}}{\omega_t^{l+1}}.
\end{align}
The probability density in momentum space then becomes
\begin{align} \label{eq: gamma FD}
\gamma^{\omega_c}_{n,m}(p)&=\frac{ n! \ }{\omega_t^{l+1} \pi \Gamma \paren{ n + l + 1}} \, p^{2\l} \, e^{-\frac{p^2}{\omega_t}} \, \paren{L_n^{\l}\paren{ \frac{p^2}{\omega_t}}}^2.
\end{align}
For the sake of simplicity, in the following the explicit dependence of the position and momentum space probability densities will be omitted.

 \begin{figure}[H]
 \begin{center}
 \begin{tabular}{cccc}
 \subfloat[]{\includegraphics[scale = 0.7]{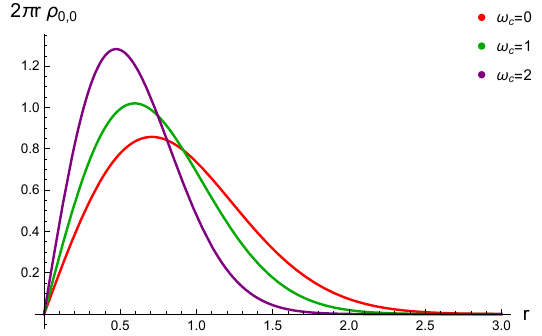}} &
 \subfloat[]{\includegraphics[scale = 0.7]{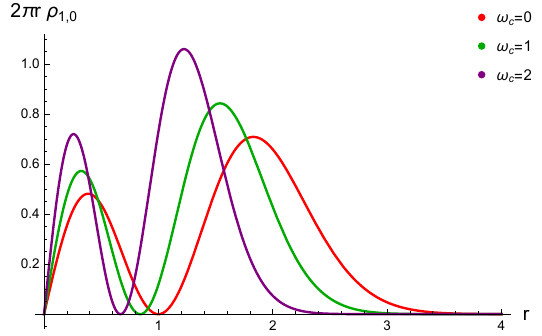}}  \\
 \subfloat[]{\includegraphics[scale = 0.7]{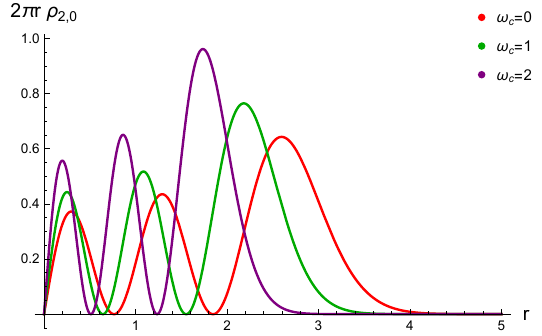}} &
 \subfloat[]{\includegraphics[scale = 0.7]{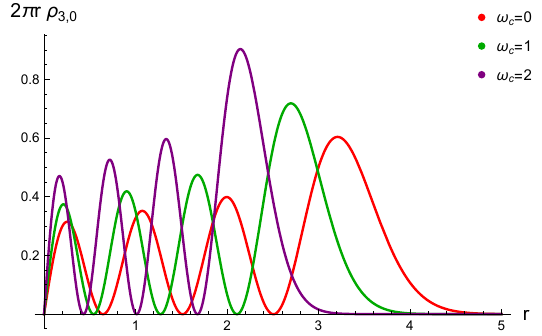}}
 \end{tabular}
 \end{center}
 \caption{Probability density of finding the particle at a distance $r$ from the origin for the Fock-Darwin system for $l=|m|=0$, $\omega=1$ and $n=0$ (A), $n=1$ (B), $n=2$ (C), and $n=3$ (D). Note that the probability density becomes more localised with increasing $\omega_c$.}
 \label{grid: density position}
 \end{figure}

 \begin{figure}[H]
  \begin{center}
 \begin{tabular}{cccc}
 \subfloat[]{\includegraphics[scale = 0.7]{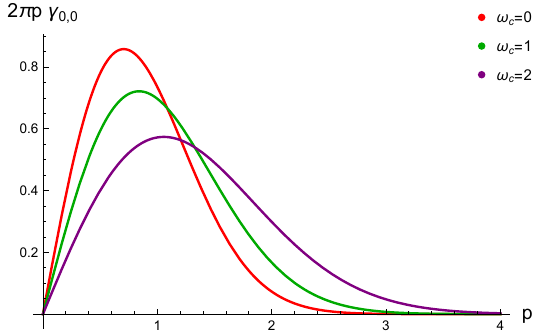}} &
 \subfloat[]{\includegraphics[scale = 0.7]{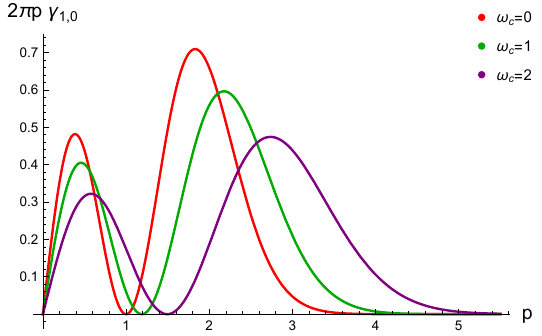}}  \\
 \subfloat[]{\includegraphics[scale = 0.7]{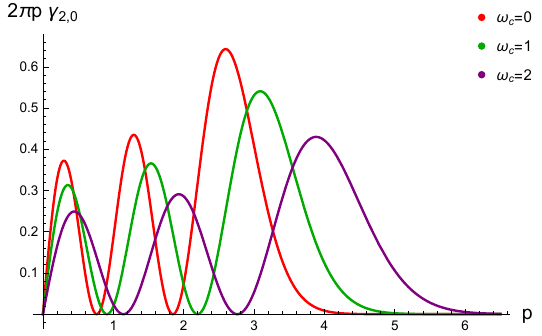}} &
 \subfloat[]{\includegraphics[scale = 0.7]{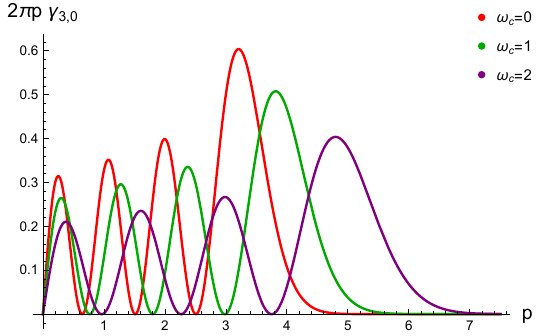}}
 \end{tabular}
  \end{center}
 \caption{Probability density of finding the particle with radial momentum $p$ in the Fock-Darwin system for $l=|m|=0$, $\omega=1$, and $n=0$ (A), $n=1$ (B), $n=2$ (C), and $n=3$ (D). Note that it becomes less localised as $\omega_c$ increases.}
 \label{grid: density momentum}
 \end{figure}

These densities for the Fock-Darwin quantum system, and therefore all entropic and dispersion measures as well, are essentially the same as those of the harmonic oscillator, but with a different frequency $\omega_t$ in (\ref{eq: omega t}). Thus, in the limit $\omega_c \to 0$, the harmonic oscillator probability densities in position and momentum spaces, respectively, are recovered
\begin{align} \label{eq: density HO 2D}
\rho_{n,m}^{HO}&=\frac{ n! \ \omega^{l+1}}{\pi \Gamma \paren{ n + l + 1}} \, r^{2\l} \, e^{-\omega r^2} \, \paren{L_n^{\l}\paren{ \omega r^2}}^2, \\ \label{eq: gamma HO 2D}
\gamma_{n,m}^{HO}&=\frac{ n! \ }{\omega^{l+1} \pi \Gamma \paren{ n + l + 1}} \, p^{2\l} \, e^{-\frac{p^2}{\omega}} \, \paren{L_n^{\l}\paren{ \frac{p^2}{\omega}}}^2.
\end{align}
Moreover, in the limit $\omega \to 0$, one recovers the Landau system. The radial probability densities reduce to
\begin{align} \label{eq: density L 2D}
\rho_{n,m}^{L}&=\frac{ n! \ \omega_c^{l+1}}{\pi \Gamma \paren{ n + l + 1}} \, r^{2\l} \, e^{-\omega_c r^2} \, \paren{L_n^{\l}\paren{ \omega_c r^2}}^2, \\ \label{eq: gamma L 2D}
\gamma_{n,m}^{L}&=\frac{ n! \ }{\omega_c^{l+1} \pi \Gamma \paren{ n + l + 1}} \, p^{2\l} \, e^{-\frac{p^2}{\omega_c}} \, \paren{L_n^{\l}\paren{ \frac{p^2}{\omega_c}}}^2.
\end{align}
Note that although the frequency changes, the mathematical structure of the densities remains identical. 


\subsection{Shannon, Rényi and Tsallis entropies of the Fock-Darwin system} \label{sec: SRT entropies}

Rényi and Tsallis entropies provide two parameter-dependent extensions of Shannon's information entropy. They are defined as functionals of the probability density $\rho(\boldsymbol{z})=\abs{\Psi(\boldsymbol{z})}^2$ associated with the quantum state $\Psi(\boldsymbol{z})$. Their definitions are
\begin{align} \label{eq: renyi entropy}
    \mathcal{R}^{(\pe)} \left[ \rho\right]&=\frac{1}{1-\pe} \loga{ \mathcal{W}^{(\pe)} \corche{\rho}}, \hspace{1cm}  \pe > 0, \ \pe \neq 1, 
 \end{align}   
 \begin{align} \label{eq: tsallis entropy}
    \mathcal{T}^{(\pe)} \left[ \rho\right]&=\frac{1}{1-\pe} \paren{ \mathcal{W}^{(\pe)} \corche{\rho}-1}, \hspace{1cm}  \pe > 0, \ \pe \neq 1,   
\end{align}
respectively. Here, $\mathcal{W}^{(\pe)} \corche{\rho}$ is the entropic moment, or frequency moment, of order $\pe$ \cite{romera2001hausdorff,jizba2016one}, which in the present two-dimensional setting, is given by
\begin{align}
     \mathcal{W}^{(\pe)} \left[\rho \right] &=\int_{\mathbb{R}^2} \rho^\pe(\boldsymbol{z})\dif \boldsymbol{z}.
\end{align}
The Shannon entropy \cite{shannon1948mathematical} is recovered from both families in the limit $\pe \to 1$:
\begin{align} \label{eq: shannon}
    \mathcal{S}\left[ \rho\right]=\lim_{\pe \to 1} \mathcal{R}^{(\pe)} \left[ \rho\right]=\lim_{\pe \to 1} \mathcal{T}^{(\pe)} \left[ \rho\right]=- \int_{\mathbb{R}^2}  \rho(\boldsymbol{z}) \loga{\rho(\boldsymbol{z})} \dif \boldsymbol{z}.
\end{align}

Although Shannon, Rényi, and Tsallis entropies are all used to quantify the delocalisation of a probability density, they do so from different perspectives. Shannon entropy corresponds to the standard additive measure and is especially relevant in conventional information-theoretic and communication contexts \cite{shannon1948mathematical}. Rényi entropy preserves additivity while introducing the parameter $\pe$, which changes the weight assigned to different parts of the distribution: large values of $\pe$ enhance the contribution of the most probable regions, whereas values below unity increase the relative importance of the tails. This flexibility makes Rényi entropy useful, for example, in multifractal problems and in systems governed by power-law structures \cite{yujun2017multiscale,chen2012renyi,bashkirov2000information}. Tsallis entropy, by contrast, is non-additive and is therefore naturally connected with situations where correlations, memory, or heavy-tailed behaviour play a central role \cite{tsallis2009nonadditive}. These quantities should therefore be regarded not as equivalent alternatives, but as complementary descriptors adapted to different statistical and physical regimes.

Entropy-based uncertainty relations are also relevant for the analysis of quantum states. In particular, for conjugated Fourier transforms in two dimensions, the Sobolev inequality can be written as \cite{olendski2019renyi}
\begin{align} \label{eq: incertidumbre sobolev}
    \paren{\frac{\pe}{\pi}}^{\frac{1}{2\pe}} \paren{\int_{\mathbb{R}^2} \rho^\pe(\boldsymbol{x})  \dif \boldsymbol{x}}^{\frac{1}{2\pe}} \geq \paren{\frac{\beta}{\pi}}^{\frac{1}{2\beta}} \paren{\int_{\mathbb{R}^2}  \gamma^\beta(\boldsymbol{p}) \dif \boldsymbol{p}}^{\frac{1}{2\beta}},\hspace{1cm} \frac{1}{\pe}+\frac{1}{\beta} = 2, \hspace{0.5cm} \frac{1}{2} <  \pe \leq 1,
\end{align}
where $\rho(\boldsymbol{x})$ and $\gamma(\boldsymbol{p})$ denote the position and momentum probability densities. By expressing this inequality in terms of the Rényi and Tsallis entropies defined above, one obtains the corresponding entropic uncertainty bounds \cite{bialynicki2006formulation,zozor2008some,rajagopal1995sobolev}. Under the same restrictions on $\pe$ and $\beta$ as above, these read
\begin{align} \label{eq: zozor N2} 
       \mathcal{R}^{(\pe)} \left[ \rho(\boldsymbol{x})\right]+ \mathcal{R}^{(\beta)} \left[ \gamma(\boldsymbol{p})\right] &\geq
       2 \log{\left( \pi \pe^\frac{1}{2\pe-2} \beta^{\frac{1}{2\beta-2}}  \right)}, \\ \label{eq: uncertainty Tsallis}
       \paren{\frac{\pe}{\pi}}^{\frac{1}{2\pe}} \paren{\paren{1-\pe}\mathcal{T}^{(\pe)}\corche{\rho(\boldsymbol{x})}+1}^\frac{1}{2\pe} &\geq \paren{\frac{\beta}{\pi}}^{\frac{1}{2\beta}} \paren{\paren{1-\beta}\mathcal{T}^{(\beta)}\corche{\gamma(\boldsymbol{p})}+1}^\frac{1}{2 \beta}.
\end{align}


\subsection{Entropic and dispersion measures of the Fock-Darwin system} \label{sec: Entropic and dispersion measures of the Fock-Darwin oscillator}

We now present the entropic moments and the Shannon, Rényi and Tsallis entropies for the Fock-Darwin system, together with some inequalities relating them. We recall that these expressions are formally similar to the ones for the harmonic oscillator with frequency $\omega$ replaced with $\omega_t$, so we refer the interested reader to the corresponding references for details. 

The Shannon entropy \cite{dehesa1997information} in position and momentum spaces are
\begin{align}
\mathcal{S}\corche{\rho_{n,m}^{\omega_c}}=\loga{2\pi}-\log \left(\frac{\omega_t 2 n!}{\Gamma\left(n+l+1\right)}\right)-\frac{n!}{\Gamma\left(n+l+1\right)}\left(J_1+J_2\right)+2 n+l+1, \\
\mathcal{S}\corche{\gamma^{\omega_c}_{n,m}}=\loga{2\pi}-\log \left(\frac{ 2 n!}{\omega_t\Gamma\left(n+l+1\right)}\right)-\frac{n!}{\Gamma\left(n+l+1\right)}\left(J_1+J_2\right)+2 n+l+1,
\end{align} 
with
\begin{align}
J_1&=\int_0^{+\infty} z^{l} e^{-z}\left(L_n^{l}(z)\right)^2 \log \left(z^l\right) \mathrm{d} z, \hspace{1cm}
J_2=\int_0^{+\infty} z^{l} e^{-z}\left(L_n^{l}(z)\right)^2 \loga{\left(L_n^{l}(z)\right)^2} \mathrm{d} z.
\end{align}

The relation of Shannon entropies of the Fock-Darwin system with the harmonic oscillator, of densities $\rho_{n,m}^{HO}$ and $\gamma_{n,m}^{HO}$, Eqs. (\ref{eq: density HO 2D}) and (\ref{eq: gamma HO 2D}) is
\begin{align} \label{eq: Shannon to Harmonic}
    \mathcal{S}\corche{\rho_{n,m}^{\omega_c}}&=-\loga{\sqrt{1+\paren{\frac{\omega_c}{\omega}}^2}}+ \mathcal{S}\corche{\rho^{HO}_{n,m}}, \\
\mathcal{S}\corche{\gamma^{\omega_c}_{n,m}}&=\loga{\sqrt{1+\paren{\frac{\omega_c}{\omega}}^2}}+ \mathcal{S}\corche{\gamma^{HO}_{n,m}}.  
\end{align}
For this reason it is easily checked that the uncertainty principle is satisfied independently from the magnetic field and values are those of the harmonic oscillator.
\begin{align} 
    \mathcal{S}\corche{\rho_{n,m}^{\omega_c}}+\mathcal{S}\corche{\gamma^{\omega_c}_{n,m}}&= \mathcal{S}\corche{\rho^{H0}_{n,m}}+\mathcal{S}\corche{\gamma^{HO}_{n,m}} \geq 2 (1+\log \pi).
\end{align}
To the best of our knowledge, the entropic moments of the $N$-dimensional harmonic oscillator have not been obtained in terms of Laguerre polynomial integrals \cite{puertas2018exact}, but rather through the $N$-fold factorisation of the one-dimensional problem. Here, we present the two-dimensional analytical expression in terms of the entropic moments of Laguerre polynomials \cite{sanchez2011direct}. A more detailed discussion of how these expressions are obtained for the FDD system can be found in subsection \ref{sec: entropic moments FDD}. The entropic moments read:
\begin{align} \label{eq: FD entropic moment}
     \mathcal{W}^{(\pe)} \corche{\rho^{\omega_c}_{n,m}} &= \frac{
     \pi^{1-\pe} (n!)^\pe (\l\pe)!
     }{
     \paren{\omega_t}^{1-\pe}
     \pe^{\l \pe+1} \Gamma \left( n + l + 1 \right)^\pe 
     }     
     \binom{n+l}{n}^{2\pe}
      F_A^{2\pe} \paren{\l \pe+1;\llave{-n}; \llave{\l+1};  \llave{\frac{1}{\pe}}}, 
      \\
    \mathcal{W}^{(\beta)} \corche{\gamma^{\omega_c}_{n,m}} &= \frac{
     \paren{\pi \omega_t}^{1-\beta} (n!)^\beta (\l\beta)!
     }{
     \beta^{\l \beta+1} \Gamma \left( n + l + 1 \right)^\beta 
     }     
     \binom{n+l}{n}^{2\beta}
      F_A^{2\beta} \paren{\l \beta+1;\llave{-n}; \llave{\l+1};  \llave{\frac{1}{\beta}}},
\end{align}
for $\alpha,\beta \in \mathbb{N}, \ \pe,\beta>1 $, where $F_A^{N}$ is the Lauricella hypergeometric series of $N$ variables \cite{abramowitz1968handbook}. The relation with the harmonic oscillator are, respectively
\begin{align} \label{eq: momento magnetico to harmonic}
    \mathcal{W}^{(\pe)} \corche{\rho^{\omega_c}_{n,m}}&=\paren{\sqrt{1+\paren{\frac{\omega_c}{\omega}}^2}}^{\pe-1} \mathcal{W}^{(\pe)} \corche{\rho^{HO}_{n,m}}, \\    \mathcal{W}^{(\beta)} \corche{\gamma^{\omega_c}_{n,m}}&=\paren{\sqrt{1+\paren{\frac{\omega_c}{\omega}}^2}}^{1-\beta} \mathcal{W}^{(\beta)} \corche{\gamma^{HO}_{n,m}}. 
\end{align}
Now, using Eq. \eqref{eq: FD entropic moment}, Rényi entropies are 
\begin{align} \label{eq:RenyiFD} \notag
      \mathcal{R}^{(\pe)} \corche{\rho^{\omega_c}_{n,m}} &=\loga{\frac{\pi}{\omega_t}} +
      \frac{\pe}{1-\pe} \loga{\frac{n!}{\Gamma\paren{n+l+1}}  \binom{n+l}{n}^2} \\&+
     \frac{1}{1-\pe} \loga{
      \frac{
     (\l\pe)!
     }{
     \pe^{\l \pe+1} 
     }
      F_A^{2\pe} \paren{\l \pe+1;\llave{-n}; \llave{\l+1};  \llave{\frac{1}{\pe}}}
      } \, ,
      \\ \notag
 \mathcal{R}^{(\beta)} \corche{\gamma^{\omega_c}_{n,m}} &=\loga{\pi \omega_t} +
      \frac{\beta}{1-\beta} \loga{\frac{n!}{\Gamma\paren{n+l+1}}  \binom{n+l}{n}^2} \\&+
     \frac{1}{1-\beta} \loga{
      \frac{
     (\l\beta)!
     }{
     \beta^{\l \beta+1} 
     }
      F_A^{2\beta} \paren{\l \beta+1;\llave{-n}; \llave{\l+1};  \llave{\frac{1}{\beta}}} \, .
      }   
\end{align}
Their relations with the ones for the harmonic oscillator are exactly the same as those for Shannon entropy
\begin{align} \label{eq: Renyi mag to Harmonic}
    \mathcal{R}^{(\pe)}\corche{\rho_{n,m}^{\omega_c}}&=-\loga{\sqrt{1+\paren{\frac{\omega_c}{\omega}}^2}}+ \mathcal{R}^{(\pe)}\corche{\rho^{HO}_{n,m}}, \\
    \mathcal{R}^{(\beta)}\corche{\gamma^{\omega_c}_{n,m}}&=\loga{\sqrt{1+\paren{\frac{\omega_c}{\omega}}^2}}+ \mathcal{R}^{(\beta)}\corche{\gamma^{HO}_{n,m}}.    
\end{align}
Similarly, from Eq. \eqref{eq: FD entropic moment}, it can be shown that Tsallis entropies are given by
\begin{align}
     \mathcal{T}^{(\pe)} \corche{\rho^{\omega_c}_{n,m}} &=\frac{1}{1-\pe} \paren{\frac{
     \paren{\pi}^{1-\pe} (n!)^\pe (\l\pe)!
     }{
     \paren{\omega_t}^{1-\pe}
     \pe^{\l \pe+1} \Gamma \left( n + l + 1 \right)^\pe 
     }     
     \binom{n+l}{n}^{2\pe}
      F_A^{2\pe} \paren{\l \pe+1;\llave{-n}; \llave{\l+1};  \llave{\frac{1}{\pe}}}-1},
      \\
  \mathcal{T}^{(\beta)} \corche{\gamma^{\omega_c}_{n,m}} &= 
  \frac{1}{1-\beta} \paren{
  \frac{
     \paren{\pi \omega_t}^{1-\beta} (n!)^\beta (\l\beta)!
     }{
     \beta^{\l \beta+1} \Gamma \left( n + l + 1 \right)^\beta 
     }     
     \binom{n+l}{n}^{2\beta}
      F_A^{2\beta} \paren{\l \beta+1;\llave{-n}; \llave{\l+1};  \llave{\frac{1}{\beta}}}-1}.
\end{align}
The relation with Tsallis entropy for the harmonic oscillator is given by
\begin{align}\label{eq:TsallisFD}
    \mathcal{T}^{(\pe)}\corche{\rho_{n,m}^{\omega_c}}&=\paren{\sqrt{1+\paren{\frac{\omega_c}{\omega}}^2}}^{1-\pe} \mathcal{T}^{(\pe)}\corche{\rho^{HO}_{n,m}} + \frac{1}{1-\pe} \paren{\paren{\sqrt{1+\paren{\frac{\omega_c}{\omega}}^2}}^{1-\pe}-1}, \\
    \mathcal{T}^{(\beta)}\corche{\gamma^{\omega_c}_{n,m}}&=\paren{\sqrt{1+\paren{\frac{\omega_c}{\omega}}^2}}^{\beta-1} \mathcal{T}^{(\pe)}\corche{\gamma^{HO}_{n,m}} + \frac{1}{1-\beta} \paren{\paren{\sqrt{1+\paren{\frac{\omega_c}{\omega}}^2}}^{\beta-1}-1} \, .
\end{align}
The uncertainty relation for the Rényi entropy (\ref{eq: zozor N2}) is also the same as the one for the harmonic oscillator.
\begin{align} 
    \mathcal{R}^{(\pe)}\corche{\rho^{\omega_c}_{n,m}}+\mathcal{R}^{(\beta)}\corche{\gamma^{\omega_c}_{n,m}}&= \mathcal{R}^{(\alpha)}\corche{\rho^{H0}_{n,m}}+\mathcal{R}^{(\beta)}\corche{\gamma^{HO}_{n,m}} \geq 2 \logacorche{
    \pi \pe^{\frac{1}{2\pe-2}}  \beta^{\frac{1}{2\beta-2}}    
    }.
\end{align}
The same happens for Tsallis entropy,
\begin{align}
    \left(\frac{\alpha}{\pi}\right)^{\tfrac{1}{\alpha}}
    \paren{(1-\alpha)\,\mathcal{T}^{(\alpha)}[\rho^{\omega_c}_{n,m}]+1}^{\tfrac{1}{2\alpha}}
    &\;\ge\;
    \left(\frac{\beta}{\pi}\right)^{\tfrac{1}{\beta}}
    \paren{(1-\beta)\,\mathcal{T}^{(\beta)}[\gamma^{\omega_c}_{n,m}]+1}^{\tfrac{1}{2\beta}}, \\
    \paren{\frac{\pe}{\pi}}^{\frac{1}{2\pe}} \corche{\mathcal{W}^{(\pe)}\corche{\rho^{\omega_c}_{n,m}}}^\frac{1}{2\pe} &\geq \paren{\frac{\beta}{\pi}}^{\frac{1}{2\beta}} \corche{\mathcal{W}^{(\pe)}\corche{\gamma^{\omega_c}_{n,m}}}^\frac{1}{2 \beta},  \\
    \paren{\frac{\pe}{\pi}}^{\frac{1}{2\pe}} \corche{\paren{\sqrt{1+\paren{\frac{\omega_c}{\omega}}^2}}^{\pe-1} \mathcal{W}^{(\pe)} \corche{\rho^{HO}_{n,m}}}^\frac{1}{2\pe} &\geq \paren{\frac{\beta}{\pi}}^{\frac{1}{2\beta}} \corche{\paren{\sqrt{1+\paren{\frac{\omega_c}{\omega}}^2}}^{1-\beta} \mathcal{W}^{(\beta)} \corche{\gamma^{HO}_{n,m}}}^\frac{1}{2 \beta}, 
\end{align}
Because $\frac{1}{\pe}+\frac{1}{\beta} = 2$, the expression simplifies to
\begin{align}
    \paren{\frac{\pe}{\pi}}^{\frac{1}{2\pe}} \corche{ \mathcal{W}^{(\pe)} \corche{\rho^{HO}_{n,m}}}^\frac{1}{2\pe} &\geq \paren{\frac{\beta}{\pi}}^{\frac{1}{2\beta}} \corche{ \mathcal{W}^{(\beta)} \corche{\gamma^{HO}_{n,m}}}^\frac{1}{2 \beta}, 
    \\
    \paren{\frac{\pe}{\pi}}^{\frac{1}{2\pe}} \paren{\paren{1-\pe}\mathcal{T}^{(\pe)}\corche{\rho_{n,m}^{HO}}+1}^\frac{1}{2\pe} &\geq \paren{\frac{\beta}{\pi}}^{\frac{1}{2\beta}} \paren{\paren{1-\beta}\mathcal{T}^{(\beta)}\corche{\gamma_{n,m}^{HO}}+1}^\frac{1}{2 \beta}.
\end{align}
\subsection{Dispersion measures for the Fock-Darwin system}
Once again we present known results for the harmonic oscillator, adapted to the Fock-Darwin system. 
The expectation values in position and momentum space \cite{dehesa2020dispersion} for the Fock-Darwin system are 
\begin{align}
    \mean{r^k}_{n,m}^{\omega_c}&=\frac{n!}{\Gamma \left( n + l + 1 \right) \paren{\omega_t}^{\frac{k}{2}}} \Gamma\paren{l+\frac{k}{2}+1} \sum_{j=0}^{n}\binom{\frac{k}{2}}{n-j}^2 \binom{\l+\frac{k}{2}+j}{j}, \\
    \mean{p^k}_{n,m}^{\omega_c}&=\frac{n!\paren{\omega_t}^{\frac{k}{2}}}{\Gamma \left( n + l + 1 \right) } \Gamma\paren{l+\frac{k}{2}+1} \sum_{j=0}^{n}\binom{\frac{k}{2}}{n-j}^2 \binom{\l+\frac{k}{2}+j}{j}.
\end{align}
The relation with the harmonic oscillator is straightforward,
\begin{align} \label{eq: rk Fock-Darwin}
   \mean{r^k}_{n,m}^{\omega_c}= \paren{1+\paren{\frac{\omega_c}{\omega}}^2}^{-\frac{k}{4}} \mean{r^k}_{n,m}^{HO}, \hspace{1cm}    \mean{p^k}_{n,m}^{\omega_c}= \paren{1+\paren{\frac{\omega_c}{\omega}}^2}^{\frac{k}{4}} \mean{p^k}_{n,m}^{HO}.
\end{align}
The uncertainty principle of the harmonic oscillator only depends on the quantum numbers $n$ and $l$, making it invariant to the frequency changes introduced by the magnetic field. For this reason, the product of the expressions in Eq. (\ref{eq: rk Fock-Darwin}) is straightforwardly independent of the magnetic field.
\begin{align}
    \mean{r^k}_{n,m}^{\omega_c} \mean{p^k}_{n,m}^{\omega_c}=\mean{r^k}_{n,m}^{HO} \mean{p^k}_{n,m}^{HO}.
\end{align}
For $k=2$, 
\begin{align} \label{eq: r2 p2 Fock-Darwin}
    \mean{r^2}_{n,m}^{\omega_c}&=\frac{1}{\omega_t}\paren{l+2n+1},
\hspace{1cm} 
\mean{p^2}_{n,m}^{\omega_c}=\omega_t\paren{l+2n+1}, 
\end{align}
and 
\begin{align}
    \mean{r^2}_{n,m}^{\omega_c} \mean{p^2}_{n,m}^{\omega_c} = \paren{l+2n+1}^2,
\end{align}
which is the known result for the harmonic oscillator \cite{dehesa2020dispersion}.

\section{The Fock-Darwin-Darboux (FDD) system} \label{sec: Magnetic Darboux III oscillator}

In this section, we firstly introduce the Darboux III surface, which is a 2-dimensional Riemannian manifold with non-constant negative curvature. Then, we describe the analogue of the previously discussed Fock-Darwin system on this surface, the Fock-Darwin-Darboux III (FDD) system, which is again an exactly solvable system. 
 
\subsection{The Darboux III surface}

Consider the 2-dimensional Riemannian manifold $\mathcal M$ with metric given by
\begin{equation}
\dif s^2 = (1+ \lambda (x^2 + y^2)) (\dif x^2 + \dif y^2)
\end{equation}
where $\{x,y\}$ are local coordinates on $\mathcal M$ and $\lambda \in \mathbb R^+$. This is a conformally flat and spherically symmetric manifold, which we call the Darboux III surface \cite{ballesteros2008maximally,ballesteros2011new,ballesteros2011quantum,darboux1915leccons}, and using polar coordinates $\{r,\theta\}$ we can rewrite the metric as 
\begin{equation}
\dif s^2 = (1+ \lambda r^2) (\dif r^2 + r^2 \dif \theta^2) .
\end{equation}
Denoting the conformal factor by $ \mu(r)=1+\lambda r^{2} $ (which is just the PDM mass function), we can finally write
\begin{align}
\dif s^{2}=\mu(r)\,\dif r^{2}+S(r)^{2}\,\dif \theta^{2},
\end{align}
where 
\begin{equation}
\label{eq:Sr}
    S(r)=r\sqrt{\mu(r)} .
\end{equation}

\begin{figure}[H]

\centering

\begin{tabular}{c c}
\multicolumn{2}{c}{\subfloat[]{\includegraphics[scale=0.8]{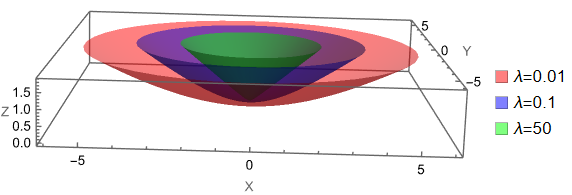}}} \\[0.5cm]
\subfloat[]{\includegraphics[scale=0.6]{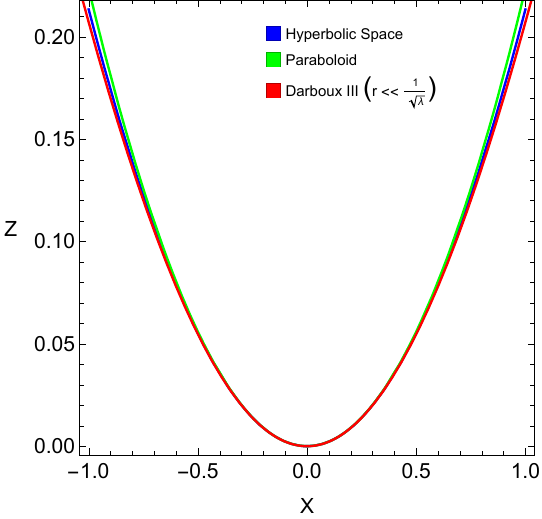}} &
\subfloat[]{\includegraphics[scale=0.6]{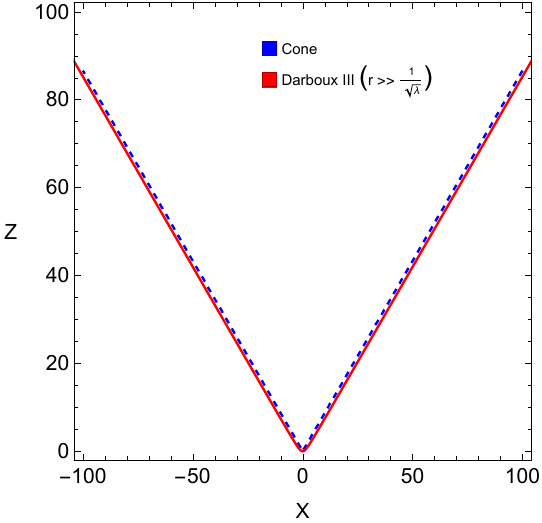}}
\end{tabular}

\caption{Embedding of the Darboux III surface in $\mathbb{R}^{2,1}$. (A) The Darboux III surface for different values of the curvature parameter $\lambda$.
(B) Intersection of the plane $Y=0$ with the Darboux III surface (with $\lambda=0.1$), the paraboloid, and hyperbolic space in a neighbourhood of $r=0$. The difference between those surfaces becomes larger as $r$ increases. It is apparent that the hyperbolic space approximates the Darboux III surface better than the paraboloid.
(C) Intersection of the plane $Y=0$ with the Darboux III surface (with $\lambda=0.1$) and the cone, for large $r$.}
\label{fig: curvatura Darboux III}

\end{figure}

The nonconstant scalar curvature of this metric, which is negative for all $r$, reads 
\begin{align} \label{eq: scalar curvature}
    R(r,\lambda)=-\frac{4 \lambda}{\paren{1+\lambda r^2}^3} .
\end{align}
For small values of $r$, the scalar curvature is
\begin{align} \label{eq: R(r,lambda) r small}
 \lim_{r \to 0}    R(r,\lambda) = -4 \lambda ,
\end{align}
while for large values of $r$ we get a flat space
\begin{equation}
\lim_{r \to \infty}  R(r,\lambda) = 0
\end{equation}
and thus we can see the Darboux III surface as a $\lambda$ deformation of the hyperbolic space going to a flat space for larger $r$. The scalar curvature is depicted in Figure \ref{fig: R(r,lambda)}, where it can be seen how it increases with $\lambda$ for sufficiently small values of $r$. We can also see that for larger $\lambda$ it tends to zero faster as $r$ increases, because the dependence on $r$ in Eq. \eqref{eq: scalar curvature} is modulated by $\lambda$. 
\begin{figure}[H]
    \centering
\includegraphics[width=0.5\linewidth]{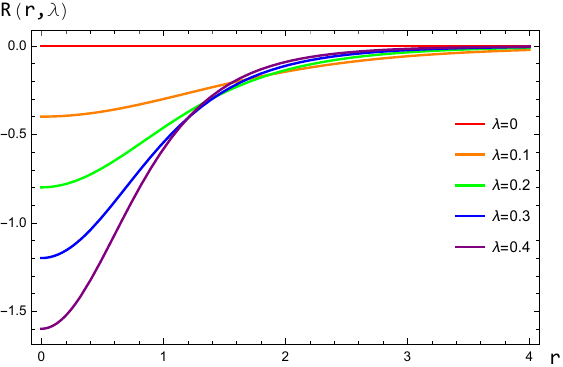}
    \caption{Scalar curvature $R(r,\lambda)$ \eqref{eq: scalar curvature} as a function of r for different values of the curvature parameter $\lambda$.}
    \label{fig: R(r,lambda)}
\end{figure}
In order to visualise the Darboux III surface, we look for an isometric embedding in $\mathbb R^{2,1}$, i.e. $\mathbb R^3$ endowed with the pseudo-Riemannian metric 
\begin{equation}
\dif s^2 = \dif X^2 + \dif Y^2 - \dif Z^2 ,
\label{eq:ambientmetric}
\end{equation}
where $\{X,Y,Z\}$ are Cartesian coordinates (it can be proved that the Darboux III surface cannot be isometrically embedded in $\mathbb R^3$ endowed with a Riemannian metric). Explicitly, this embedding is given by the surface of revolution
\begin{equation}
\begin{split}
    X(r, \theta) &= S(r) \cos(\theta) \\
    Y(r, \theta) &= S(r) \sin(\theta) \\
    Z(r, \theta) &= h(r)\\
\end{split}
\label{eq:embedding}
\end{equation}
where $S(r)$ is given by \eqref{eq:Sr} and $h(r)$ is the solution of the following differential equation
\begin{align}
\left(\frac{dh}{dr}\right)^{2}=-\mu(r)+\left(\frac{dS}{dr}\right)^{2},
\label{eq:eqdiffhr}
\end{align}
which reads 
\begin{align}
h(r)=\frac{3\sqrt{\paren{\paren{1+\lambda r^{2}}\paren{2+3\lambda r^{2}}}}-\sqrt{3}\,\operatorname{arcsinh}\!\paren{\sqrt{\paren{2+3\lambda r^{2}}}}}{6\sqrt{\lambda}} + C ,
\end{align}
where $C$ is and arbitrary constant (fixing a value of $C$ amounts to perform an isometry of the ambient space, and here we fix $C$ such that $h(0)=0$). Figure \ref{fig: curvatura Darboux III} (A) shows the Darboux III surface embedded in $\mathbb R^{2,1}$ for different values of $\lambda$.

We can get a deeper insight into the geometry of the Darboux III surface by looking at the neighbourhood of $r = 0$ and analysing it for large $r$. First, at the neighbourhood of $r = 0$ (with $r \ll \dfrac{1}{\sqrt{\lambda}}$), we can make a Taylor series expansion of the r.h.s. of Eq. \eqref{eq:eqdiffhr}. We have that $S(r) \approx r + \dfrac{\lambda}{2} r^3$ and thus Eq. \eqref{eq:eqdiffhr} reads 
\begin{equation}
    \left(\frac{dh}{dr}\right)^{2}=2 \lambda r^2 ,
\end{equation}
and straightforwardly its positive solution reads
\begin{equation}
    h(r) = \sqrt{\frac{\lambda}{2}} r^2 + C .
\end{equation}
Therefore, at the neighbourhood of $r = 0$, the Darboux III surface can be approximated at leading order by the elliptic paraboloid with equation 
\begin{equation}
    Z= \sqrt{\frac{\lambda}{2}} (X^2 + Y^2) .
\end{equation}
The scalar curvature of this elliptic paraboloid at $r=0$ (recall that the ambient space is endowed with the pseudo-Riemannian metric \eqref{eq:ambientmetric}), which coincides with the limit $r \to 0$ of the scalar curvature of the Darboux III surface, see Eq. \eqref{eq: R(r,lambda) r small}) is $R = -4 \lambda$. In fact, as shown in Figure \ref{fig: curvatura Darboux III} (B), the hyperbolic space of constant scalar curvature $R = -4 \lambda$, given by the implicit equation 
\begin{equation}
    X^2 + Y^2 - Z^2 = - \frac{1}{2 \lambda} \, ,
    \label{eq:hyperbolicspace}
\end{equation}
is indeed a better approximation for the Darboux III surface. This can be checked by parametrizing all three spaces using the coordinates defined by \eqref{eq:embedding}, and considering the Taylor expansions of their heigh functions. For the Darboux III surface we obtain
\begin{equation}
    h(r) \approx \sqrt{\frac{\lambda}{2}} r^2 (1 + \frac{\lambda}{8} r^2) ,
\end{equation}
for the hyperbolic space we get
\begin{equation}
    h(r) \approx \sqrt{\frac{\lambda}{2}} r^2 (1 + \frac{\lambda}{2} r^2) ,
\end{equation}
and finally for the paraboloid
\begin{equation}
    h(r) \approx \sqrt{\frac{\lambda}{2}} r^2 (1 + \lambda r^2) .
\end{equation}
From these expressions, it is apparent that while the first-order approximations for these three surfaces of revolution are equal, at second order the Darboux III surface is more similar to the hyperbolic space. 

For sufficiently large $r$ ($r \gg \dfrac{1}{\sqrt{\lambda}}$), we have that $\mu(r) \approx \lambda r^2$ and $S(r) \approx \sqrt \lambda r^2$ and thus Eq. \eqref{eq:eqdiffhr} takes the simpler form
\begin{equation}
    \left(\frac{dh}{dr}\right)^{2}=3 \lambda r^2
\end{equation}
and its positive solution reads
\begin{equation}
    h(r) = \frac{\sqrt{3 \lambda}}{2} r^2 + C ,
\end{equation}
and thus the limit of the Darboux III surface for large $r$ is given by the cone with equation
\begin{equation}
    X^2 + Y^2 - \left( \frac{2Z}{\sqrt {3}} \right)^2 = 0 ,
\end{equation}
which interestingly does not depend on $\lambda$ (see Figure \ref{fig: curvatura Darboux III} C). The previous study shows that the Darboux III surface is a conformally flat surface that can be thought of as a deformation of the hyperbolic space that tends to a cone for large $r$.

\subsection{Hamiltonian and probability density of the FDD system}

We now consider the FDD system describing the motion of a charged particle moving on the Darboux III surface subjected to a magnetic field. We recall that, in the absence of the magnetic field, this is the 2-dimensional version of the so-called Darboux III oscillator, which is an exactly solvable and maximally superintegrable (in general $N$-dimensional) quantum nonlinear oscillator defined on a radially symmetric space with nonconstant negative curvature \cite{ballesteros2011quantum,ballesteros2023shannon}. 

Using the Schrödinger quantization \cite{ballesteros2011quantum}, the Hamiltonian of the FDD system is given by \cite{ballesteros2024dunkl}
\begin{align}
    \mathcal{H}=\frac{1}{2\paren{1+\lambda (\hat{x}^2+\hat{y}^2)}} \paren{\paren{\hat{p}_x + \frac{e B}{2 c} \hat{y}}^2+ \paren{\hat{p}_y - \frac{e B}{2 c} \hat{x}}^2+\frac{\omega^2}{2} \paren{\hat{x}^2
    +\hat{y}^2} }
\label{eq:FDDHamiltonian}
\end{align}
where, as in the Fock-Darwin system, $\omega_c=\frac{e B}{2 c}$ is the Larmor frequency and $\hat{L}_z=\hat{x} \hat{p}_y-\hat{y} \hat{p}_x$ is the angular momentum operator. By introducing once again the modulation frequency as in Eq.~(\ref{eq: omega t}), $\omega_t=\sqrt{\omega_c^2+\omega^2}$, the Hamiltonian becomes
\begin{align}
\mathcal{H}=\frac{1}{2 \paren{1+\lambda \hat{r}^2}} \paren{\left(\hat{p}_x^2+\hat{p}_y^2\right)+\omega_t^2 \hat{r}^2-2 \omega_c \hat{L}_z},
\end{align}
with $\hat{r}^2=\hat{x}^2+\hat{y}^2$. The eigenvalues and eigenfunctions of this quantum nonlinear oscillator have been computed in \cite{ballesteros2024dunkl}, and we refer the interested reader to \cite{ballesteros2011quantum,ballesteros2024dunkl} for details. In particular, the discrete energy levels of the FDD system are given by
\begin{align} \label{eq: En Darboux magnetico}
    E_{n,m}^{\lambda,\omega_c}
    &= -\hbar m \omega_c
    - \hbar^2 \lambda (2 n + |m| + 1)^2
    + \hbar \sqrt{
        \paren{
            \hbar \lambda (2 n + |m| + 1)^2
            + m \omega_c
        }^2
        + \omega_t^2 (2 n + |m| + 1)^2
        - m^2 \omega_c^2 \, ,
    }
\end{align}
where, in the same manner as in the FD system, $n=0,1,2.\dots$ and $m=0,\pm 1,\pm 2,\dots$, and the associated wave functions are formally the same as those of the Darboux III oscillator with an effective energy-dependent frequency $\Ot$ 
given by
\begin{align} \label{eq: Ot}
\Ot&=\sqrt{\omega^2_t-2 \lambda E_{n,m}^{\lambda,\omega_c} }.
\end{align}
Explicitly,
\begin{equation} \label{eq: Psi nl lambda}
\Psi_{n,m}^{\lambda,\omega_c}(r,\varphi)
= \mathcal{N}^{\lambda,\omega_c}_{n,m}\,
\sqrt{1 + \lambda r^2}\,
r^l\,
e^{-\frac{(\beta_t^\lambda)^2}{2} r^2}\,
L_n^{l}\!\left( (\beta_t^\lambda)^2 r^2 \right)\,
\frac{ e^{i m \varphi} }{\sqrt{2\pi}},
\hspace{1cm}
\beta_t^\lambda = \sqrt{\frac{ \Omega_{n,m}^{\lambda,\omega_c} }{\hbar}} \,,
\end{equation}
with $l=|m|$, and where the radial normalization constant is given by
\begin{equation} \label{eq: norm darboux magnetico}
\normNla =  \beta_t^{l+1}\sqrt{\frac{2 n!}{\Gamma \left( n + l + 1 \right)}} \sqrt{\frac{1}{1 + \left( 2n + l + 1 \right) \frac{\lambda}{\Omega_{n,m}^{\lambda,\omega_c}}}}. 
\end{equation}

 \begin{figure}[H]
  \begin{center}
 \begin{tabular}{cccc}
 \subfloat[]{\includegraphics[scale = 0.9]{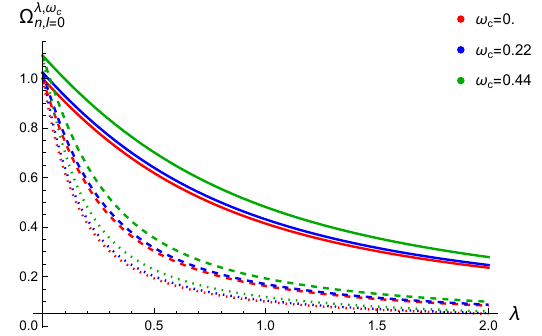}} &
 \subfloat[]{\includegraphics[scale = 0.9]{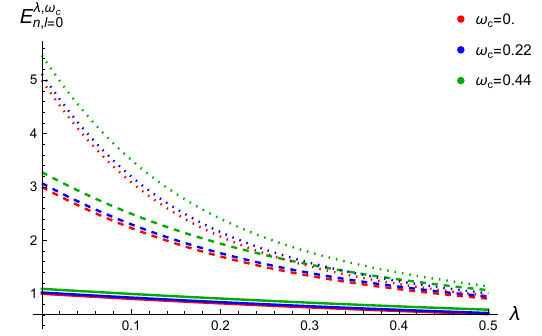}}
 \end{tabular}
  \end{center}
 \caption{(A) Effective frequency $\Ot$ \eqref{eq: Ot} and (B) energy $E_{n,m}^{\lambda,\omega_c}$ \eqref{eq: En Darboux magnetico} of the FDD system as a function of $\lambda$ for different values of $\omega_c$, with $n=0$ (straight lines), $n=1$ (dashed lines) and $n=2$ (dotted lines). In all the cases $l=0$, $\omega=1$.}
 \label{grid: omega energy FDD}
 \end{figure}

For the sake of clarity, from now on, we take units such that $\hbar=1$. The probability density in position space of the FDD eigenfunctions is given by 
\begin{equation} \label{eq: density magnetic darboux}
\rho_{n,m}^{\lambda,\omega_c}(r,\varphi) = \frac{\paren{\mathcal{N}^{\lambda,\omega_c}_{n,m}}^2}{2 \pi} \paren{1 + \lambda r^2} \, r^{2l} \, e^{-\Ot r^2} \, \paren{L_n^{l} \left( \Ot r^2 \right)}^2.
\end{equation}

\begin{figure}[H]
  \begin{center}
 \begin{tabular}{cccc}
 \subfloat[]{\includegraphics[scale = 0.9]{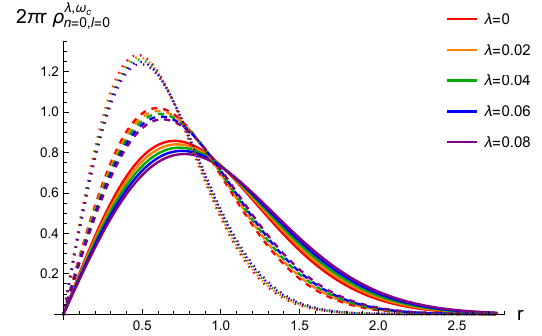}} &
 \subfloat[]{\includegraphics[scale = 0.9]{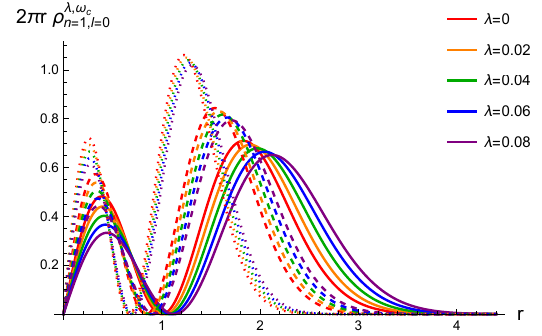}}
 \end{tabular}
  \end{center}
 \caption{FDD probability density in position space of finding the particle at a distance $r$ from the origin for different values of $\lambda$ and with $\omega_c=0$ (straight lines), $\omega_c=1$ (dashed lines) and $\omega_c=2$ (dotted lines). (A) $n=0$, (B) $n=1$. In all the cases $l=0$, $\omega=1$.}
 \label{grid: density FDD position}
 \end{figure}

Note that the effective frequency $\Ot$~\eqref{eq: En Darboux magnetico} decreases with $n$ and $\lambda$, but increases with $\omega_c$, whereas the energy increases with $n$ and $\omega_c$ and decreases with $\lambda$, as seen in Figure \ref{grid: omega energy FDD}. This affects the probability densities, as shown in Figure \ref{grid: density FDD position}: they delocalise as $\lambda$ increases, but localise as $\omega_c$ increases. All of this anticipates the opposing behaviour between the curvature effects of the FDD oscillator and the strength of the magnetic field, which will be studied more in detail in the following Section.

\begin{figure}[H]
\begin{tabular}{cccc}
\subfloat[]{\includegraphics[scale = 0.8]{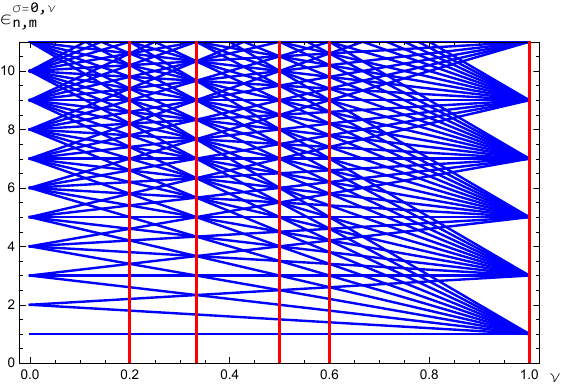}} &
\subfloat[]{\includegraphics[scale = 0.8]{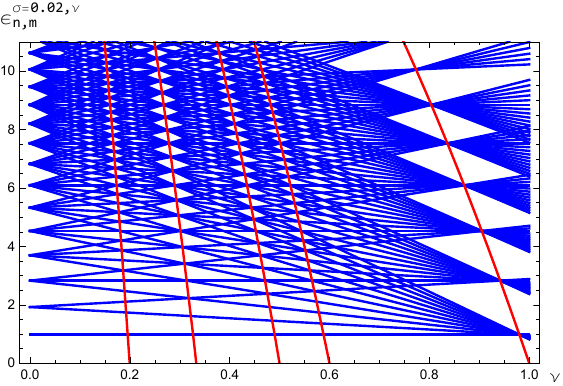}}
\end{tabular}
\caption{Dimensionless energy for the Fock-Darwin (A) and FDD with $\sigma=0.02$ (B) as a function of $\nu$. The vertical lines in (A) correspond to the rational values of $\nu$ given by $\frac{q_1}{q_2}=\tfrac{1}{5}, \tfrac{1}{3}, \tfrac{1}{2}, \tfrac{3}{5}, 1$. The parabolic lines in (B) correspond to the plot of Eq. (\ref{eq: nu lambda}) for the same values of $\tfrac{q_1}{q_2}$.}
\label{fig: dimensionless energy}
\end{figure}

It is interesting to note that the energy spectrum is degenerate for certain values of the quantum numbers and of the curvature parameter $\lambda$. We define the FDD dimensionless energy by
\begin{align} \label{eq: energia reducida Darboux III magnetico}
\epsilon_{n,m}^{\sigma,\nu}=\frac{E_{n,m}^{\lambda,\omega_c}}{\omega_t}=\sqrt{
\paren{ \sigma c^2 + \nu m }^2
+ c^2
- \nu^2 m^2
}
- \sigma c^2
- \nu m,
\end{align}
where we have written equation \eqref{eq: c}, with $c=2n+\abs{m}+1$, once again
\begin{align}
    \sigma=\frac{\lambda}{\omega_t}\, ,
\end{align}
and the value of $\nu$ is $\omega_c/\omega_t$ (as defined in Eq. \ref{eq: nu}). Therefore, degeneracy is obtained when $\epsilon_{n_1,m_1}^{\sigma,\nu}=\epsilon_{n_2,m_2}^{\sigma,\nu}$, which results in the analytical expression 
\begin{align} \label{eq: nu solucion}
\nu=\frac{
\sigma (c_1 - c_2)(c_1 m_2 - c_2 m_1)
+ \abs{c_1 - c_2}\,
\sqrt{
\sigma^2 \paren{c_2 m_1 - c_1 m_2}^2 + \paren{m_1-m_2}^2
}
}{
(m_1 - m_2)^2.
}
\end{align}
As seen in Figure~\ref{fig: dimensionless energy}, the nonvanishing parameter $\sigma$ distorts the Fock–Darwin spectrum, although for sufficiently small values of $\sigma$, the energy levels are only slightly shifted with respect to the standard Fock–Darwin solutions, as expected. However, as $\sigma$ increases, these deviations become more pronounced, thus making the degeneracy structure to arise when different, albeit close, values for $\nu$ are considered. For this reason, $\sigma=0.02$ was chosen in Figure~\ref{fig: dimensionless energy} (B), where in the limit $\sigma \to 0$, which is shown in Figure~\ref{fig: dimensionless energy} (A), we recover
\begin{align}
    \nu=\frac{\abs{c_1-c_2}}{\abs{m_1-m_2}} = \frac{q_1}{q_2}, 
\end{align}
with $q_1 \in \mathbb Z$, $q_2 \in \mathbb Z^*$ such that $|q_1| \leq |q_2|$ and $q_1 q_2 \geq 0$, which is similar to the Fock-Darwin case (see Eq. \eqref{eq: nu FD}). 

For the FDD oscillator case in Figure \ref{fig: dimensionless energy} (B) all the intersections obey Eq. (\ref{eq: nu solucion}) and are contained in curves, whose equations can be explicitly obtained. We define the dimensionless $\lambda-$deformed frequency as
\begin{align}
    \nu_{\lambda}=\frac{\omega_c}{\Omega_{n,m}^{\lambda,\omega_c}}=\frac{\omega_c}{\sqrt{\omega_t^2
    -2 \lambda E_{n,m}^{\lambda,\omega_c}}}=\frac{\nu}{\sqrt{1
    -2 \sigma \epsilon_{n,m}^{\sigma,\nu}}}.
\end{align}

By taking squares and inverting the equation, we obtain that for a specific $\nu_{\lambda}=\tfrac{q_1}{q_2} \in \mathbb Q$, the equation of the degeneracy curves is 
\begin{align} \label{eq: nu lambda}
    \epsilon_{n,m}^{\sigma,\nu}=\frac{1}{2\sigma} \paren{1-\frac{\nu^2}{q_1^2/q_2^2}} = \frac{1}{2\sigma} \paren{1-\frac{\omega_c^2}{(q_1^2/q_2^2)(\omega^2 + \omega_c^2)}}.
\end{align}

\begin{figure}[H]
\begin{tabular}{cccc}
\subfloat[]{\includegraphics[scale = 0.8]{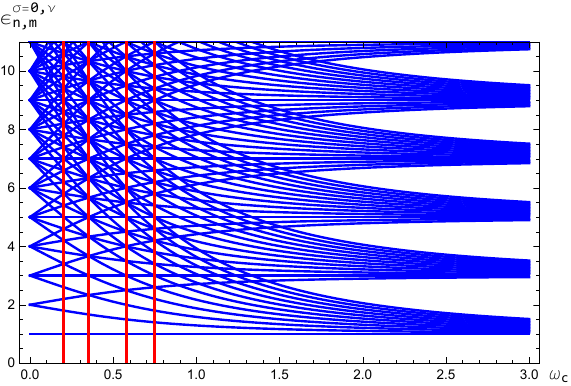}} &
\subfloat[]{\includegraphics[scale = 0.8]{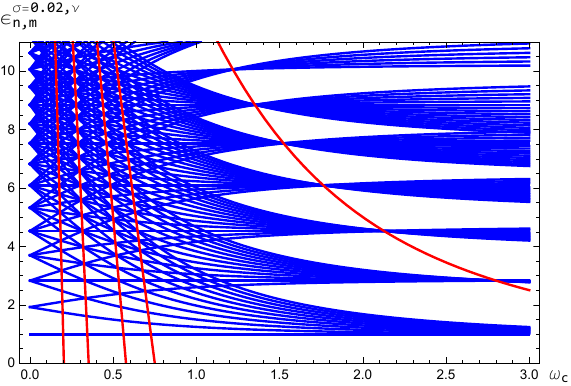}}
\end{tabular}
\caption{Dimensionless energy for the Fock-Darwin (A) and the FDD with $\sigma=0.02$ (B) as a function of $\omega_c$ (with $\omega=1$). The vertical lines in (A) correspond to $\frac{q_1}{q_2}=\tfrac{1}{5}, \tfrac{1}{3}, \tfrac{1}{2}, \tfrac{3}{5}$. The parabolic lines in (B) correspond to the plot of Eq. (\ref{eq: nu lambda}) for $\frac{q_1}{q_2}=\tfrac{1}{5}, \tfrac{1}{3}, \tfrac{1}{2}, \tfrac{3}{5}, 1$.}
\label{fig: dimensionless energy vs omegac}
\end{figure}

Figure \ref{fig: dimensionless energy vs omegac} shows the energy spectrum and degeneracy curves of the Fock-Darwin (A) and FDD (B) oscillators as a function of the Larmor frequency $\omega_c$.

It is worth stressing that the previous discussion shows that the degeneracy properties of the eigenvalues of the Landau-Darboux system, which is obtained as the $\omega \to 0$ limit of the FDD Hamiltonian \eqref{eq:FDDHamiltonian},
namely
\begin{align}
    \mathcal{H}_{LD}=\frac{1}{2\paren{1+\lambda (\hat{x}^2+\hat{y}^2)}} \paren{\paren{\hat{p}_x + \frac{e B}{2 c} \hat{y}}^2+ \paren{\hat{p}_y - \frac{e B}{2 c} \hat{x}}^2} \, ,
    \label{LD}
\end{align}
are completely different to the ones of the standard Landau system 
\begin{align}
    \mathcal{H}_{L}=\frac{1}{2}\paren{\paren{\hat{p}_x + \frac{e B}{2 c} \hat{y}}^2+ \paren{\hat{p}_y - \frac{e B}{2 c} \hat{x}}^2} \, ,
\end{align}
which is obtained when $\lambda$ also vanishes. In fact, the usual Landau system on the plane corresponds to $\nu=1$, and in that case all eigenvalues have infinite degeneracy (Landau levels), as it is shown in Figure \ref{fig: dimensionless energy} (A). In contradistinction, when $\lambda\neq 0$ the spectrum for $\nu=1$  shows no degeneracy at all, and thus Landau levels disappear on the curved space model. As shown in Figure \ref{fig: dimensionless energy}  (B), in order to reconstruct one of the energy levels with infinite degeneracy one has to perform a fine tuning by making use of lower value for $\nu$, which physically implies the introduction of a curved oscillator potential on the Darboux III surface through a nonvanishing $\omega$ parameter. Moreover, for a fixed $\lambda$ each level with infinite degeneracy has to be obtained with a different curved oscillator frequency $\omega$, which means that the Landau-Darboux system for a given $\lambda$ can have at most one Landau level. This total degeneracy breaking of the Landau system induced by the curvature of the underlying space is the main feature of the Landau-Darboux Hamiltonian~\eqref{LD}.


\section{Entropies and dispersion measures for the FDD system} \label{sec: resultados FDD}

In this Section we compute the Shannon, Rényi, and Tsallis entropies analytically in position space (starting from the probability density of the FDD oscillator given in Eq. \eqref{eq: density magnetic darboux}) and numerically in momentum space, and we study the possible effects of the magnetic field on them. Finally, dispersion measures in both spaces, along with the uncertainty principle, are also analytically obtained and discussed.

\subsection{Shannon, Rényi and Tsallis entropies for the FDD system}

Regarding the Shannon, Rényi and Tsallis entropies, we are able to provide explicit analytical expressions for them on position space. In momentum space, the nonlinear nature of the underlying Darboux III surface precludes the existence of a closed-form expression for the Fourier transform of the wave-function, and thus for the probability density in momentum space. Therefore, the study in momentum space of the relevant quantities will be performed numerically. 

\subsubsection{Shannon information entropy}

The Shannon entropy for the Darboux III oscillator has been recently obtained in \cite{ballesteros2023shannon}. For the FDD system in position space, similar expressions are obtained just substituting the oscillator frequency by the effective frequency $\Ot$. Doing this, we have that the Shannon information entropy in position space for the FDD nonlinear oscillator is given by
\begin{align} 
\mathcal{S}_\rho^{n, m, \lambda} &= -\log \Ot - \log \left( \frac{2 n!}{\Gamma\left(n+l+1\right)} \right) + \log \left( 1 + (2n+l+1) \frac{\lambda}{\Ot} \right) \\ \notag
&\quad - \frac{n!}{\Gamma\left(n+l+1\right)} \frac{1}{1 + (2n+l+1) \frac{\lambda}{\Ot}} 
\left( J_1 + \frac{\lambda}{\Ot} \tilde{J}_1 + J_2 + \frac{\lambda}{\Ot} \tilde{J}_2 
+ J_3^{\frac{\lambda}{\Ot}} + \frac{\lambda}{\Ot} \tilde{J}_3^{\frac{\lambda}{\Ot}} \right) \\ \notag
&\quad + \frac{1}{1 + (2n+l+1) \frac{\lambda}{\Ot}} 
\left( (2n+l+1) + \frac{\lambda}{\Ot} \left( 6n^2 + 6ln + 6n + l + l^2 + 2l + 2 \right) \right)-J_Y,
\end{align}
where
\begin{align}
    J_Y=-\int_0^{2\pi} \abs{Y_l(\varphi)}^2 \loga{\abs{Y_l(\varphi)}^2} d\varphi=-\int_0^{2\pi} \frac{1}{2\pi} \log{\frac{1}{2\pi}} d\varphi = \log{\frac{1}{2\pi}},
\end{align}
and 
\begin{align}
J_1&=\int_0^{+\infty} z^{l} e^{-z}\left(L_n^{l}(z)\right)^2 \log \left(z^l\right) \,\mathrm{d} z, 
\hspace{1cm} \tilde{J}_1=\int_0^{+\infty} z^{l+1} e^{-z}\left(L_n^{l}(z)\right)^2 \log \left(z^l\right) \,\mathrm{d} z, 
\\
J_2&=2 \int_0^{\infty} z^{l} e^{-z}\left(L_n^{l}(z)\right)^2 \loga{L_n^{l}(z)} \,\mathrm{d} z, 
\hspace{0.5cm}
\tilde{J}_2=2\int_0^{\infty} z^{l+1} e^{-z}\left(L_n^{l}(z)\right)^2 \loga{L_n^{l}(z)} \,\mathrm{d} z, 
\\
J_3^\alpha&=\int_0^{+\infty} z^{l} e^{-z}\left(L_n^{l}(z)\right)^2 \log (1+\alpha z) \,\mathrm{d} z, 
\hspace{0.5cm}
\tilde{J}_3^\alpha=\int_0^{+\infty} z^{l+1} e^{-z}\left(L_n^{l}(z)\right)^2 \log (1+\alpha z) \,\mathrm{d} z.
\end{align}

\subsubsection{Entropic moments} \label{sec: entropic moments FDD}

Entropic moments are needed to obtain the Rényi and Tsallis entropies, so we compute them first. Since the wave function factorises in its radial and angular parts the entropic moments also factorize, namely
\begin{align}
    \mathcal{W}^{(\pe)} \left[\rho_{n,m}^{\lambda,\omega_c} \right]=\mathcal{W}^{(\pe)} \left[\abs{R_{n,m}^{\lambda,\omega_c}}^2 \right] \mathcal{W}^{(\pe)} \left[\abs{Y_{m}}^2 \right].
\end{align}
As we have already seen, the radial symmetry of the system induces a radial symmetry of the wave-function, and thus the angular part is given by the usual spherical harmonics. Therefore the angular entropic moment is $\lambda$-invariant, and simply reads
\begin{align}
    \mathcal{W}^{(\pe)} \left[\abs{Y_{m}}^2 \right]=\paren{\frac{1}{2 \pi}}^\pe \int_0^{2\pi} \abs{\expo{i \l \varphi}}^{2\pe} \dif\varphi =\paren{2\pi}^{1-\pe} .
\end{align}
However, the radial entropic moment differs from the Fock-Darwin case, namely
\begin{align}
     \mathcal{W}^{(\pe)} \left[\abs{R_{n,m}^{\lambda,\omega_c}}^2 \right] =\paren{\mathcal{N}^{\lambda,\omega_c}_{n,m}}^{2\pe} \int_0^\infty (1+\lambda r^2)^\pe \paren{L_n^\l(\Ot r^2)}^{2\pe} r^{2\l\pe} \expo{-\pe \Ot r^2} r \dr.
\end{align}
Making the change of variable $\Ot \pe r^2 = t$ and using the binomial theorem, we obtain
\begin{align} \label{eq: momento entropico integral con cambio de variable}
     \mathcal{W}^{(\pe)} \left[\abs{R^{\lambda,\omega_c}_{n,m}}^2 \right] =\paren{\mathcal{N}^{\lambda,\omega_c}_{n,m}}^{2\pe} \sum_{k=0}^\pe \binom{\pe}{k} \frac{\lambda^k}{2 \paren{\Ot \pe}^{l \pe+k+1}} \int_0^\infty  \paren{L_n^\l \paren{\frac{t}{\pe}}}^{2\pe} t^{\l\pe+k} \expo{-t} \dt.
\end{align}
Entropic moments for Laguerre polynomials have been derived in \cite{sanchez2011direct} . It was achieved by using the linearization formula of Srivastava–Niukkanen \cite{srivastava2003some}
\begin{align} \label{eq: linearization Laguerre}
    x^\mu L_{n_1}^{l_1}(t_1 x) \cdots L_{n_r}^{l_r}(t_r x)=\sum_{k=0}^\infty \Theta_k\paren{\mu,\beta,r,\llave{n_i},\llave{l_i},\llave{t_i} }L_k^{\beta}(x),
\end{align}
where the coefficients are
\begin{align} \notag
    \Theta_k(&\mu,\beta,r,\llave{n_i},\llave{l_i},\llave{t_i}=\paren{\beta+1}_\mu \binom{n_1+l_1}{n_1} \cdots \binom{n_r+l_r}{n_r} \times 
    \\ \label{eq: theta coeff}
    &F_A^{r+1} \corche{\beta+\mu+1;-m1,\dots,-m_r,-k;\l_1+1,\dots \l_r+1,\beta+1; t_1,\dots t_r,1},
\end{align}
and $F_A^{r+1}$ is the Lauricella hypergeometric series of $r+1$ variables \cite{abramowitz1968handbook}, which in general is defined as
\begin{align} \label{eq:  lauricella def}
    F_A^{N}(a,\llave{b_i},\llave{c_i};\llave{x_i})=\sum_{i_1,\dots,i_N=0}^\infty \frac{(a)_{i_1+\cdots i_N} (b_1)_{i_1} \cdots (b_N)_{i_N}}
    {
    (c_1)_{i_1} \cdots (c_N)_{i_N} i_1! \cdots i_N!
    } x_1^{i_1} \cdots x_N^{i_N}.
\end{align}
Using the former for integral in Eq. \eqref{eq: momento entropico integral con cambio de variable}, we obtain
\begin{align} \notag
     \mathcal{W}^{(\pe)} \left[\abs{R^{\lambda,\omega_c}_{n,m}}^2 \right] &=\frac{1}{2 \pi} \paren{\mathcal{N}^{\lambda,\omega_c}_{n,m}}^{2\pe} \binom{n+l}{n}^{2\pe} \\
     &\sum_{k=0}^\pe \binom{\pe}{k} \frac{\lambda^k}{2 \paren{\Ot \pe}^{l \pe+k+1}} \paren{l \pe+k}!  F_A^{2\pe} \corche{\l \pe+k+1;\llave{-n}; \llave{\l+1};  \llave{\frac{1}{\pe}}}.
\end{align}
Finally, substituting the normalization constant, the entropic moment reads, for $\pe \in \mathbb{N}\setminus\{1\},$
\begin{align} \label{eq: entropic moment magnetic darboux}
    \mathcal{W}^{(\pe)} \left[\rho_{n,m}^{\lambda,\omega_c}\right]=\paren{\frac{\pi}{\Ot}}^{1-\pe}  \paren{\frac{n!}{\Gamma \left( n + l + 1 \right)} \, \, \frac{1}{1 + \left( 2n + l + 1 \right) \frac{\lambda}{\Omega_{n,m}^{\lambda,\omega_c}}}}^{\pe} \binom{n+l}{n}^{2\pe}  \eta_{n,m}^{\lambda,\omega_c},
\end{align}
with 
\begin{align} \label{eta magnetico}
    \eta_{n,m}^{\lambda,\omega_c}=\sum_{k=0}^\pe \binom{\pe}{k} \paren{\frac{\lambda}{\Ot \pe}}^k \frac{\paren{l \pe+k}!}{\pe^{l \pe +1}}  \, F_A^{2\pe} \corche{\l \pe+k+1;\llave{-n}; \llave{\l+1};  \llave{\frac{1}{\pe}}}.
\end{align}
In the limit $\lambda \to 0$, the only surviving term in the sum above corresponds to $k=0$. This can be reasoned due to the fact that the $k$ index comes from the binomial theorem of the term $\paren{1+\lambda r^2}^\pe$ that reduces to $1$ when $\lambda \to 0$, so 
\begin{align}
    \lim_{\lambda \to 0} \eta_{n,m}^{\lambda,\omega_c}=\frac{\paren{l \pe}!}{\pe^{l \pe +1}}  \, F_A^{2\pe} \corche{\l \pe+1;\llave{-n}; \llave{\l+1};  \llave{\frac{1}{\pe}}}.
\end{align}
Since $\lim_{\lambda \to 0} \Otnl = \omega_t$ and the term $\frac{1}{1 + \left( 2n + l + 1 \right) \frac{\lambda}{\Omega_{n,m}^{\lambda,\omega_c}}}$ goes to $1$, the entropic moment then reduces to
\begin{align}
    \lim_{\lambda \to 0} \mathcal{W}^{(\pe)} \left[\rho_{n,m}^{\lambda,\omega_c}\right]=\paren{\frac{\pi}{\omega_t}}^{1-\pe}  \paren{\frac{n!}{\Gamma \left( n + l + 1 \right)}}^{\pe} 
     \binom{n+l}{n}^{2\pe}
      F_A^{2\pe} \paren{\l \pe+1;\llave{-n}; \llave{\l+1};  \llave{\frac{1}{\pe}}}, 
\end{align}
which is the entropic moment of the FD system in Eq. \eqref{eq: FD entropic moment}.

\subsubsection{Rényi and Tsallis entropies}

Once the entropic moments in Eq.~\eqref{eq: entropic moment magnetic darboux} have been explicitly computed, Rényi and Tsallis entropies are easily obtained, and they are given respectively by
\begin{align} \notag
     \mathcal{R}^{(\pe)} \left[\rho_{n,m}^{\lambda,\omega_c}\right]&=\loga{\frac{\pi}{\Ot}}+\frac{\pe}{1-\pe} \loga{\frac{n!}{\Gamma \left( n + l + 1 \right)} \binom{n+l}{n}^{2}
      } \\\label{eq: renyi magnetic darboux}    &+ \loga{\frac{1}{1 + \left( 2n + l + 1 \right) \frac{\lambda}{\Omega_{n,m}^{\lambda,\omega_c}}}} + \frac{1}{1-\pe} \loga{\eta_{n,m}^{\lambda,\omega_c}},
\end{align}
and
\begin{align} \label{eq: tsallis magnetic darboux}
    \mathcal{T}^{(\pe)} \left[\rho_{n,m}^{\lambda,\omega_c}\right]= \frac{1}{1-\pe} \paren{\paren{\frac{\pi}{\Ot}}^{1-\pe}  \paren{\frac{n!}{\Gamma \left( n + l + 1 \right)} \, \, \frac{1}{1 + \left( 2n + l + 1 \right) \frac{\lambda}{\Omega_{n,m}^{\lambda,\omega_c}}}}^{\pe} \binom{n+l}{n}^{2\pe}  \eta_{n,m}^{\lambda,\omega_c}-1},
\end{align}
for $\pe \in \mathbb{N}\setminus\{1\}$. Once again, in the limit $\lambda \to 0$, we recover the expressions \eqref{eq:RenyiFD} and \eqref{eq:TsallisFD} for the Fock–Darwin Rényi and Tsallis entropies, respectively. 

Note the dependence of the entropies on the magnetic field and on the curvature parameter $\lambda$. To study this behaviour, we recall that the effective frequency $\Ot$ increases with $\omega_c$, but decreases with $\lambda$, as shown in Figure \ref{grid: omega energy FDD}. The parametric dependence of the entropic moments $\mathcal{W}^{(\pe)} \left[\rho_{n,m}^{\lambda,\omega_c}\right]$ on $\lambda$ and $\omega_c$ is given by
\begin{align}
    \mathcal{W}^{(\pe)} \left[\rho_{n,m}^{\lambda,\omega_c}\right] \propto \frac{\paren{\Ot}^{\pe-1}}{\paren{1+(2n+l+1)\frac{\lambda} {\Ot}}^\pe} \, \eta_{n,m}^{\lambda,\omega_c}. 
\end{align}
We analyse their behaviour explicitly in the following two regimes:

\paragraph{Regime of large $\frac{\lambda}{\Ot}$:}
This case corresponds to large values of $\lambda$ and/or small values of $\omega_c$. The constant $\eta_{n,m}^{\lambda,\omega_c}$ contains several terms, but the dominant contribution for large $\frac{\lambda}{\Ot}$ is the one with $k=\pe$, namely $\paren{\frac{\lambda}{\Ot}}^\pe$. Therefore,
\begin{align}
    \mathcal{W}^{(\pe)} \left[\rho_{n,m}^{\lambda,\omega_c}\right] \sim \frac{\paren{\Ot}^{2\pe-1}}{\lambda^\pe} \,  \paren{\frac{\lambda}{\Ot}}^\pe = \paren{\Ot}^{\pe-1},
\end{align}
which is small in this regime. Moreover, the behaviour of the entropic moments is the same as $\Ot$: it increases with $\omega_c$ and decreases with $\lambda$.

\paragraph{Regime of small $\frac{\lambda}{\Ot}$:}
In this case, the term with $k=0$ is the largest contribution to the constant $\eta_{n,m} ^{\lambda,\omega_c}$, so that it loses its dependence on the parameters. Moreover, the denominator tends to one, and therefore
\begin{align}
    \mathcal{W}^{(\pe)} \left[\rho_{n,m}^{\lambda,\omega_c}\right] \sim \paren{\Ot}^{\pe-1},
\end{align}
again. However, in this regime $\Ot$ is large because $\lambda \ll \omega_c$. Thus, the entropic moments again become larger for stronger magnetic fields or smaller values of the curvature parameter. 

\begin{figure}[H]
\begin{tabular}{cccc}
\subfloat[]{\includegraphics[scale = 0.7]{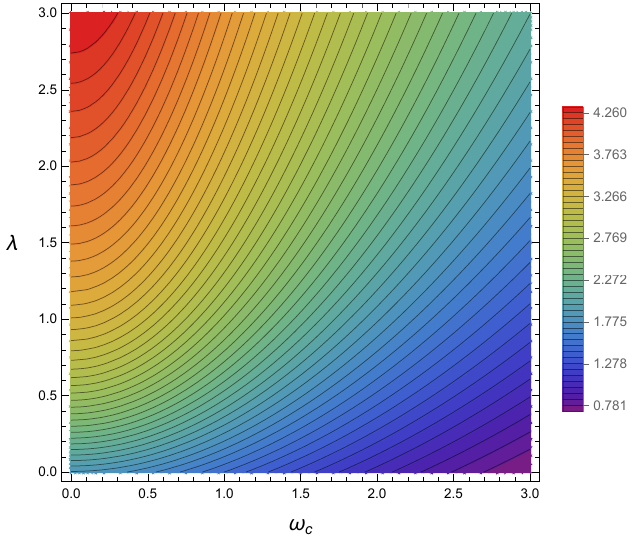}} &
\subfloat[]{\includegraphics[scale = 0.7]{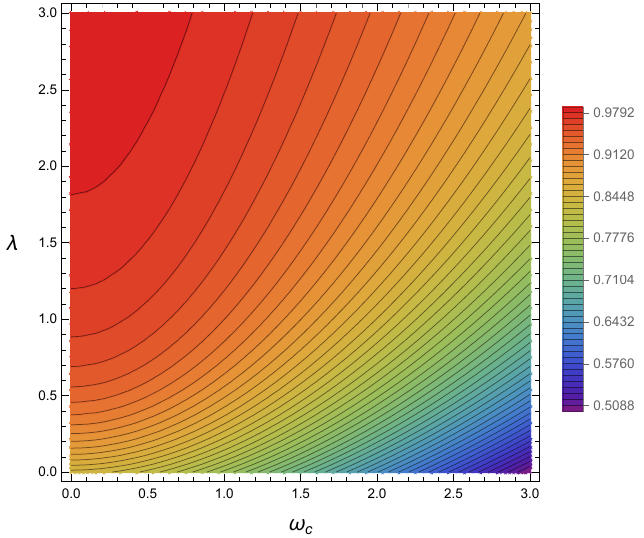}}
\end{tabular}
\caption{Rényi \eqref{eq: renyi magnetic darboux}(A) and Tsallis \eqref{eq: tsallis magnetic darboux} (B) entropies for the ground state $(n=l=0)$ as a function of $\lambda$ and $\omega_c$ (with $\omega=1$ and $\pe=2$). Entropies increase with $\lambda$ but decrease with $\omega_c$. The same behaviour is found for other values of $n$ and $l$.}
\label{grid: entropias analiticas magneticas darboux}
\end{figure}

Because of the factor $\frac{1}{1-\pe}$ appearing in the construction of the Rényi and Tsallis entropies, Eqs. \eqref{eq: renyi entropy} and \eqref{eq: tsallis entropy}, an increase in the entropic moment leads to a decrease in the Rényi and Tsallis entropies, and vice versa. Therefore, this analytical analysis shows that the Rényi and Tsallis entropies increase in position space with $\lambda$ but decrease with $\omega_c$, for both regimes above. This behaviour is shown in Figure \ref{grid: entropias analiticas magneticas darboux}, where it can be appreciated that the behaviour of these quantities in the intermediate regime follows exactly the same trends.

To obtain these expressions in momentum space, we compute the Fourier transform
\begin{align}
    \tilde \Psi(\textbf{p})&=\mathcal{F}_2 \llave{\Psi_{n,m}^{\lambda,\omega_c}(x)}=\int_{\mathbb{R}^2} \expo{-i \textbf{r} \cdot \textbf{p}} \Psi(\textbf{r}) d\textbf{r}
\end{align}
in polar coordinates. Using the Hankel transform \cite{baddour2011two} we can write
\begin{align}
    \tilde \Psi(\textbf{p})&=
   \frac{1}{2\pi}
     \int_0^\infty \int_0^{2 \pi}  \psi(r,\varphi) r \dr \dif \varphi
=\frac{1}{i^l} \int_0^\infty f_k(r) J_k\paren{r \ p} r \dr ,
\end{align}
where $p$ is the conjugate coordinate of $r$ and
\begin{align}
    f_k(r)=\frac{1}{2 \pi} \int_0^{2 \pi} \Psi \paren{r,\varphi} \expo{-i k \varphi} d\varphi.
\end{align}
In this case, 
\begin{align}
    f_k(r)=\frac{1}{2 \pi} \int_0^{2 \pi} R_{n,m}^{\lambda,\omega_c}(r) \Ye_m(\varphi) \expo{-i k \varphi} d\varphi=R_{n,m}^{\lambda,\omega_c}(r) \delta_{km},
\end{align}
with $\delta_{km}$ being the Kronecker delta, and thus
\begin{align} \label{eq: fourier trans cyl}
\tilde{\Psi}_{n,m}^{\lambda,\omega_c}(p)=\frac{1}{i^l} \int_0^\infty R_{n,m}^{\lambda,\omega_c}(r) \delta_{km} J_k\paren{r \ p} r \dr =\frac{1}{i^l} \int_0^\infty R_{n,m}^{\lambda,\omega_c}(r) J_m\paren{r \ p} r \dr.
\end{align}
Therefore, the probability density in momentum space is given by
\begin{align} \label{eq: density p cyl space}
\gamma_{n,m}^{\lambda,\omega_c}(p)=\abs{\tilde{\Psi}_{n,m}^{\lambda,\omega_c}(p)}^2=\abs{\int_0^\infty R_{n,m}^{\lambda,\omega_c}(r) J_m\paren{r \ p} r \dr}^2.
\end{align}

\begin{figure}[H]
  \begin{center}
 \begin{tabular}{cccc}
 \subfloat[]{\includegraphics[scale = 0.9]{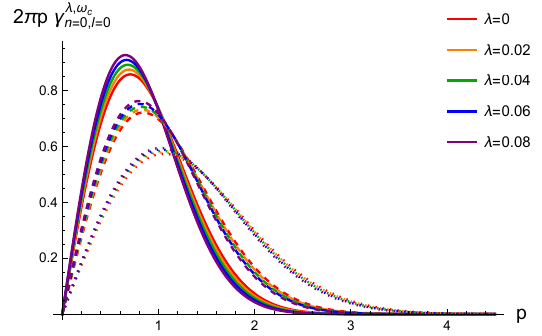}} &
 \subfloat[]{\includegraphics[scale = 0.9]{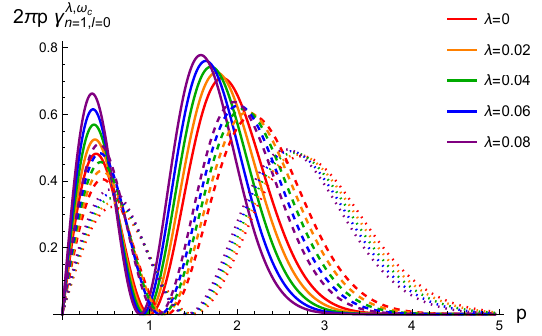}}
 \end{tabular}
  \end{center}
 \caption{FDD probability density of finding the particle with a radial momentum $p$ for different values of $\lambda$ and with $\omega_c=0$ (straight lines), $\omega_c=1$ (dashed lines) and $\omega_c=2$ (dotted lines). (A) $n=0$, (B) $n=1$. In all the cases $l=0$, $\omega=1$.}
 \label{grid: density FDD momentum}
 \end{figure}

Since the integral above does not have a known closed-form analytical solution, we are not able to provide analytical expressions for entropic moments (and neither for Rényi and Tsallis entropies) and therefore to compute these quantities we rely on numerical analysis. The probability density in momentum space, as shown in Figure \ref{grid: density FDD momentum}, varies with $\lambda$ and $\omega_c$ in the opposite way to position space: it localises as $\lambda$ increases and delocalises as $\omega_c$ increases.
Having computed the numerical Fourier transform, we compute, again numerically,  the entropies of the FDD system in momentum space.

\begin{figure}[H]
\begin{tabular}{cccc}
\subfloat[]{\includegraphics[scale = 0.9]{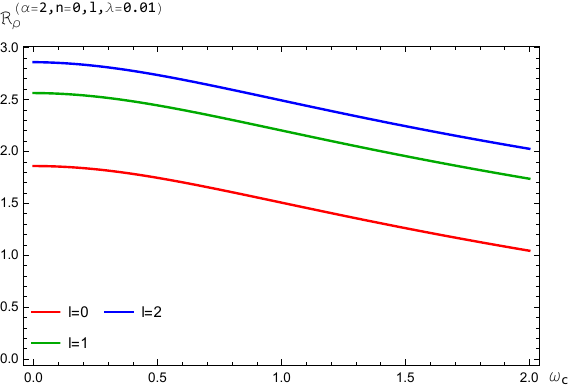}} &
\subfloat[]{\includegraphics[scale = 0.9]{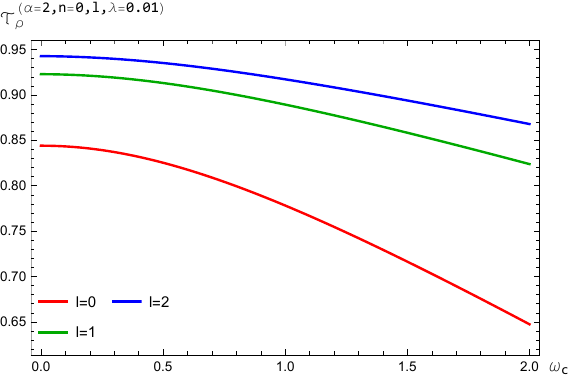}}
\end{tabular}
\caption{Rényi (A) and Tsallis (B) entropies in position space as a function of $\omega_c$ for $m=l=0,1,2$. In all cases $n=0$, $\pe=2$, $\omega=1$ and $\lambda=0.01$.}
\label{grid: entropias posicion distinta l}
\end{figure}

\begin{figure}[H]
\begin{tabular}{cccc}
\subfloat[]{\includegraphics[scale = 0.9]{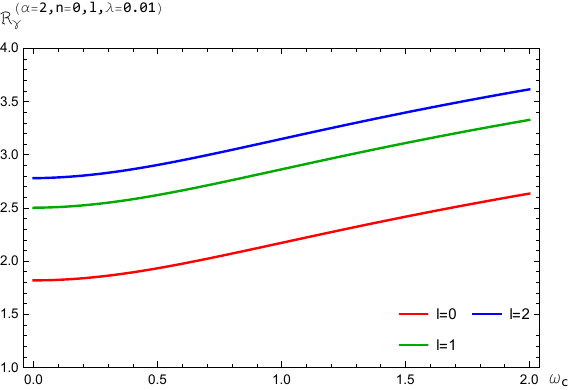}} &
\subfloat[]{\includegraphics[scale = 0.9]{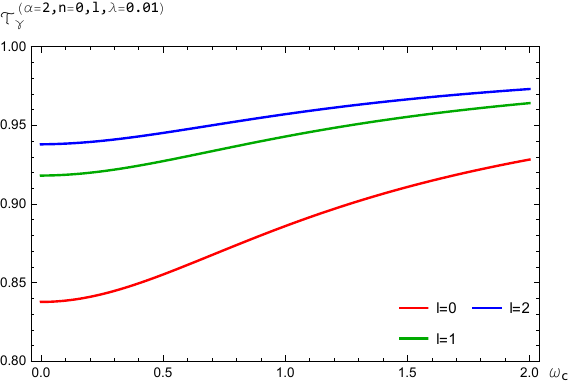}}
\end{tabular}
\caption{Rényi (A) and Tsallis (B) entropies in momentum space as a function of $\omega_c$ for $m=l=0,1,2$. In all cases $n=0$, $\pe=2$, $\omega=1$ and $\lambda=0.01$.}
\label{grid: entropias momento distinta l}
\end{figure}

The following figures show the most remarkable properties of these quantities and their dependence with the different parameters of the system. Interestingly, when increasing the value of $l$, in Figures \ref{grid: entropias posicion distinta l} and \ref{grid: entropias momento distinta l}, entropies increase in both position and momentum space. In other words, when increasing the magnetic field with a higher value of the angular momentum, entropies increase to a higher value in both spaces. No new qualitative behaviour arises when further increasing $l$. For this reason, most of the entropic analysis is focused on the case $l=0$.

\begin{figure}[H]
\begin{tabular}{cccc}
\subfloat[]{\includegraphics[scale = 0.9]{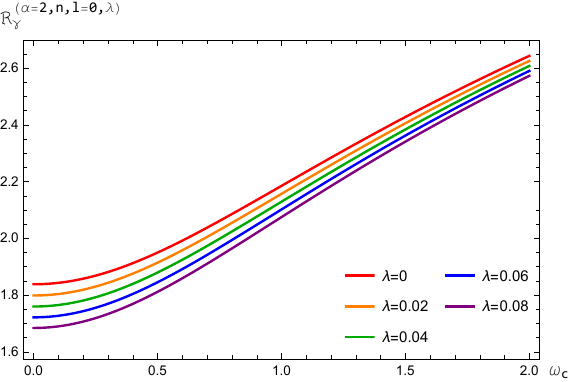}} &
\subfloat[]{\includegraphics[scale = 0.9]{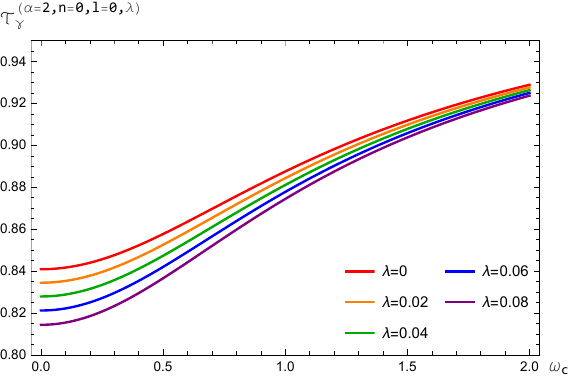}} \\
\subfloat[]{\includegraphics[scale = 0.9]{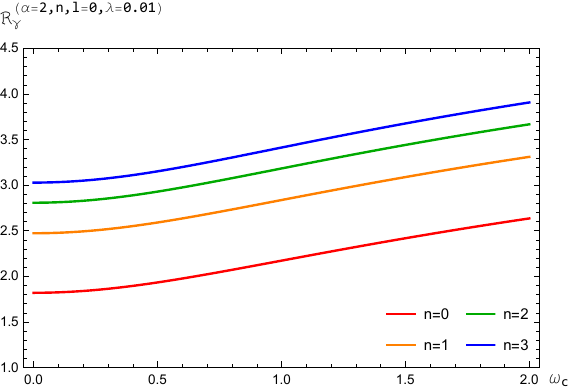}} &
\subfloat[]{\includegraphics[scale = 0.9]{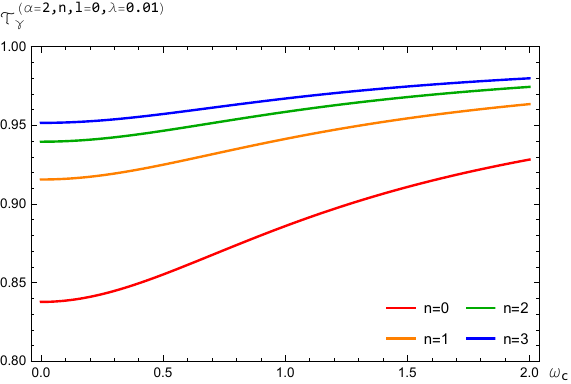}}
\end{tabular}
\caption{Rényi (A,C) and Tsallis (B,D) entropies in momentum space as a function of $\omega_c$. (A,B) $n=0$ and different values of $\lambda$ and (C,D) $\lambda=0.01$ and different values of $n$. Entropies decrease with $\lambda$ but increase with $n$ and $\omega_c$. In all cases $l=0$, $\omega=1$, $\pe=2$.}
\label{grid: entropias numericas magneticas darboux}
\end{figure}

Entropies in momentum space increase with $\omega_c$, but decrease with $\lambda$, as shown in Figure \ref{grid: entropias numericas magneticas darboux} (A,B). This is the opposite of the behaviour found in position space and can be explained by examining the localisation of the probability densities in Figure \ref{grid: density FDD momentum}. We can also observe in Figure \ref{grid: entropias numericas magneticas darboux} (C,D) that the entropies increase with $n$, as as in position space.

Thus, this analysis has shown that the curvature and the magnetic field have opposite effects: the curvature tends to spread the state in configuration space, whereas the magnetic coupling effect is to increase the effective confinement.

\subsubsection{Entropy-based uncertainty relations}

As described in the Introduction, Rényi and Tsallis information entropies satisfy generalized uncertainty relations. In order to study how these uncertainty relations depend on the parameters of the FDD system, we define the uncertainty functions
\begin{align} \label{eq: xi magnetico N2} 
       \xi \corche{\mathcal{R}^{(\pe,n,l,\lambda)}} 
       &= \mathcal{R}^{(\pe)} \corche{\rho_{n,m}^{\lambda,\omega_c}} 
       + \mathcal{R}^{(\beta)} \corche{\gamma_{n,m}^{\lambda,\omega_c}} 
       - 2 \log{\left( \pi \pe^\frac{1}{2\pe-2} \beta^{\frac{1}{2\beta-2}}  \right)}, \\
       \xi \corche{\mathcal{T}^{(\pe,n,l,\lambda)}} 
       &= \paren{\frac{\pe}{\pi}}^{\frac{1}{2\pe}} 
          \paren{ \paren{1-\pe}\,\mathcal{T}^{(\pe)}\corche{\rho_{n,m}^{\lambda,\omega_c}}+1 }^{\tfrac{1}{2\pe}}
       - \paren{\frac{\beta}{\pi}}^{\frac{1}{2\beta}} 
          \paren{ \paren{1-\beta}\,\mathcal{T}^{(\beta)}\corche{\gamma_{n,m}^{\lambda,\omega_c}}+1 }^{\tfrac{1}{2\beta}} ,
\end{align}
where $\frac{1}{\alpha}+\frac{1}{\beta}=2$ and $\frac{1}{2}<\alpha<1$. 

\begin{figure}[H]
\begin{tabular}{cccc}
\subfloat[]{\includegraphics[scale = 0.8]{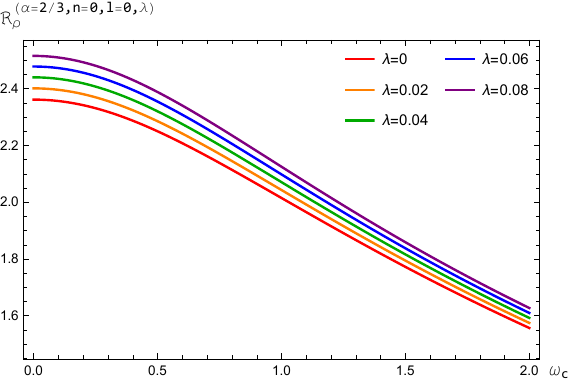}} &
\subfloat[]{\includegraphics[scale = 0.8]{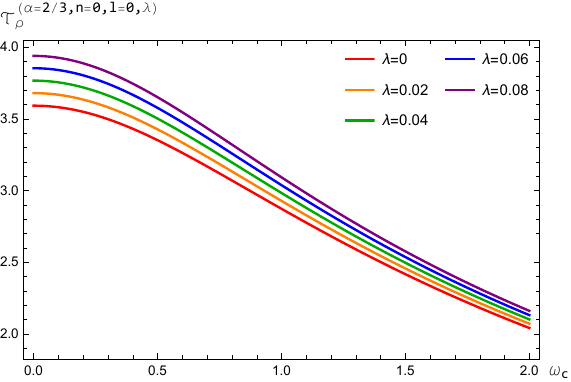}} \\
\subfloat[]{\includegraphics[scale = 0.8]{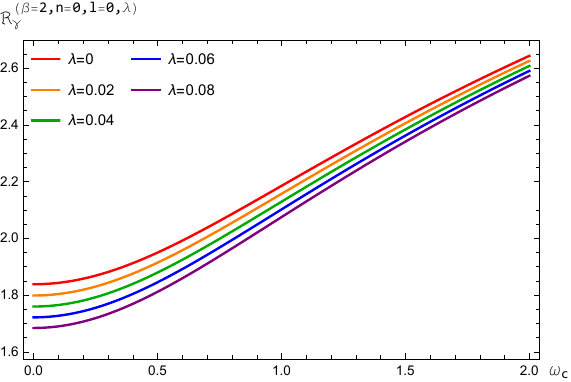}} &
\subfloat[]{\includegraphics[scale = 0.8]{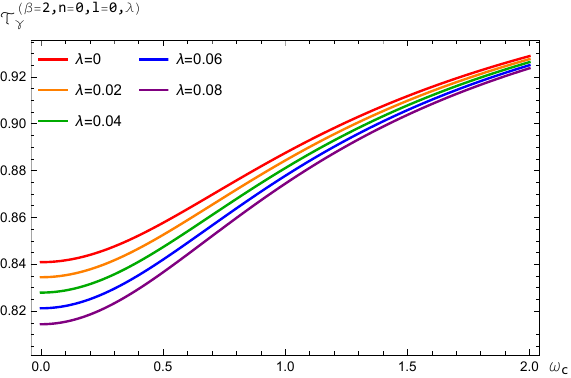}} \\
\subfloat[]{\includegraphics[scale = 0.8]{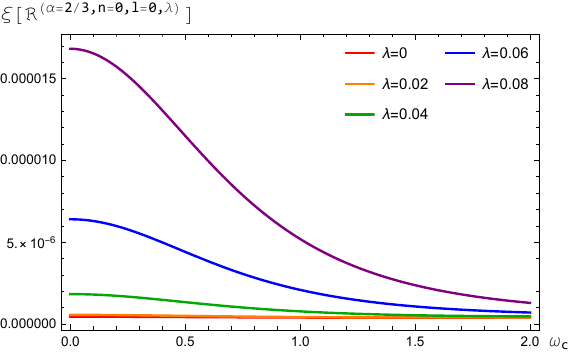}} &
\subfloat[]{\includegraphics[scale = 0.8]{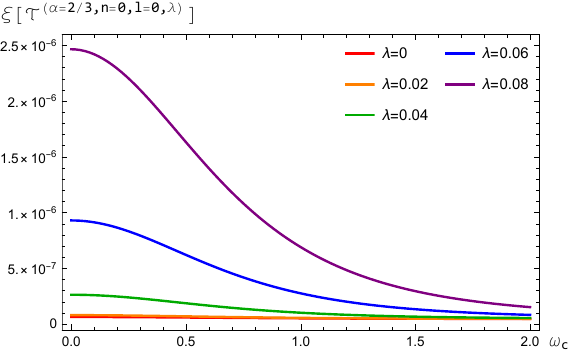}}
\end{tabular}
\caption{Rényi (left column) and Tsallis (right column) entropies in position space (1st row), in momentum space (2nd row) and uncertainty function $\xi$ (3rd row) as functions of $\omega_c$ and different values of $\lambda$. In all cases $n=l=0$, $\omega=1$, $\pe=2/3$ and $\beta=2$.}
\label{grid: entropias incertidumbre}
\end{figure}

It is well-known that the ground state of the harmonic oscillator saturates the uncertainty principle for Rényi and Tsallis entropies, and thus the uncertainty function vanishes in this case \cite{puertas2018exact}. Therefore, since the Fock–Darwin oscillator uncertainty relations are the same as those of the harmonic oscillator, the same is true for the ground state of the Fock–Darwin system ($n=l=\lambda=0$, see Section \ref{sec: Entropic and dispersion measures of the Fock-Darwin oscillator}). However, for the FDD system the uncertainty functions not only depend on the curvature parameter $\lambda$ but also on the magnetic field, in such a way that the effects of $\lambda$ and the magnetic field are opposite, similarly to the case of the entropies. In addition, a large magnetic field makes the system to behave more like the Fock–Darwin oscillator and therefore the uncertainty functions tend to those from the harmonic oscillator (zero for the ground state) for any value of $\lambda$, provided that $\omega_c$ is sufficiently large. In Figure \ref{grid: entropias incertidumbre} the behaviour here described can be clearly appreciated for the ground state (for any other state the qualitative behaviour is similar).

\subsection{Dispersion measures of the FDD system}

The goal of this section is to determine the analytical values of $\mean{r^2}_{n,m}^{\lambda,\omega_c}$ and $\mean{p^2}_{n,m}^{\lambda,\omega_c}$ and to analyze the effects of the magnetic field on the uncertainty principle relating these two quantities. In general, for central potentials it was shown in \cite{dehesa2020dispersion} that
\begin{align}
\mean{r^2}_{n,m}^{\lambda,\omega_c} \mean{p^2}_{n,m}^{\lambda,\omega_c} \geq \paren{\frac{N}{2}+l}^2.
\end{align}
In our case, we have a two-dimensional system, $N=2$, and thus
\begin{align}
\mean{r^2}_{n,m}^{\lambda,\omega_c} \mean{p^2}_{n,m}^{\lambda,\omega_c} \geq \paren{l+1}^2,
\label{eq:dispersion_UP}
\end{align}
which, for $l=0$, results in
\begin{align}
\mean{r^2}_{n,m=0}^{\lambda,\omega_c} \mean{p^2}_{n,m=0}^{\lambda,\omega_c} \geq 1.
\label{eq:dispersion_UP0}
\end{align}

\subsubsection{Dispersion measures in position space}
We start by recalling that mean values for Laguerre polynomials have been derived in\cite{sanchez2011direct} by using the integral
\begin{align} \label{eq: laguerre expected}
    \mean{t^s}_{L_{n,m}^{\pe,\beta}}=\int_0^{\infty} t^s \mathrm{e}^{-t} L_n^\alpha(x) L_m^\beta(t) \mathrm{d} t=\Gamma(s+1) \sum_{r=0}^{\min (n, m)}(-1)^{n+m}\binom{s-\alpha}{n-r}\binom{s-\beta}{m-r}\binom{s+r}{r} ,
\end{align}
which for $n=m$ and $\pe=\beta$ takes the simpler form
\begin{align} \label{eq: mean xs laguerre} 
\mean{t^s}_{L_{n}^{\pe}}=\int_0^{\infty} x^s \mathrm{e}^{-x} \paren{L_n^\alpha(x)}^2  \mathrm{d} x=\Gamma(s+1) \sum_{j=0}^{n}\binom{s-\alpha}{n-j}^2 \binom{s+j}{j} .
\end{align}
Applying this result to the FDD system, we obtain
\begin{align}
    \mean{r^k}_{n,m}^{\lambda,\omega_c}&=\frac{1}{2\pi}\int_{0}^{2\pi}\int_0^\infty \paren{\mathcal{N}^{\lambda,\omega_c}_{n,m}}^2 \, \paren{1+\lambda r^2} r^{2\l+k} \, e^{-\Ot r^2} \, \paren{L_n^{\l}\paren{ \Ot r^2}}^2  r \dr \dif \varphi \\
    &=\paren{\mathcal{N}^{\lambda,\omega_c}_{n,m}}^2 \int_0^\infty  \, \paren{1+\lambda x^2}r^{2\l+k} \, e^{-\Ot r^2} \, \paren{L_n^{\l}\paren{ \Ot r^2}}^2  r \dr ,
\end{align}
and performing the change of variable $\Ot r^2=t$, we can separate the previous integral in two parts as follows
\begin{align}
     \mean{r^k}_{n,m}^{\lambda,\omega_c}&=\frac{\paren{\mathcal{N}_{n,m}^{\lambda,\omega_c}}^2 }{2 \paren{\Ot}^{l+\frac{k}{2}+1}} \paren{
    \int_0^\infty  \, t^{\l+\frac{k}{2}} \, e^{-t} \, \paren{L_n^{\l}\paren{t}}^2  \dt + \frac{\lambda}{\Ot}\int_0^\infty  \, t^{\l+\frac{k+2}{2}} \, e^{-t} \, \paren{L_n^{\l}\paren{t}}^2  \dt  } .
\end{align}
Finally, using again the integral (\ref{eq: mean xs laguerre}) for $s=l+\frac{k}{2}$ and $s=l+\frac{k+2}{2}$ and substituting the normalization constant, (\ref{eq: norm darboux magnetico}) the radial expectation value for the FDD oscillator is given by
\begin{align} \notag
    \mean{r^k}_{n,m}^{\lambda,\omega_c}&=\frac{1}{\paren{\Ot}^{\frac{k}{2}}+(2n+l+1) \lambda \paren{\Ot}^{\frac{k}{2}-1}} 
    \frac{n! \ \G{l+\frac{k}{2}+1}}{(n+l)!} 
    \\ \label{eq: rk darboux}
    &\times
    \sum_{j=0}^n \binom{\frac{k}{2}}{n-j}^2 \binom{l+\frac{k}{2}+j}{j} +\frac{\lambda}{\Ot} \paren{l+\frac{k}{2}+2} \sum_{j=0}^n \binom{\frac{k}{2}+1}{n-j}^2 \binom{l+\frac{k}{2}+j+1}{j} .
\end{align}
It can be easily checked that for $\lambda=0$, the last term vanishes and one recovers the Fock-Darwin mean value given by Eq. \ref{eq: rk Fock-Darwin}. Additionally, for $k=2$, we get
\begin{align} \label{eq: r2 darboux}
    \mean{r^2}_{n,m}^{\lambda,\omega_c}=\frac{1}{\Ot+(2n+l+1) \lambda} \paren{2n+l+1 +\frac{\lambda}{\Ot} \paren{l^2+l (6 n+3)+6 n (n+1)+2}}
\end{align}
which is once again easily reduced for $\lambda=0$ to the Fock-Darwin expression given by Eq. \ref{eq: r2 p2 Fock-Darwin} . 
For the ground state of the FDD oscillator this expression is further reduced and reads
\begin{align} \label{eq: r2 ground}
    \mean{r^2}_{0,0}^{\lambda,\omega_c}=\frac{2\lambda+\Otoo}{\lambda \Otoo+\paren{\Otoo}^2} .
\end{align}

\begin{figure}[H]
\centering
\begin{tabular}{cccc}
\subfloat[]{\includegraphics[scale = 0.9]{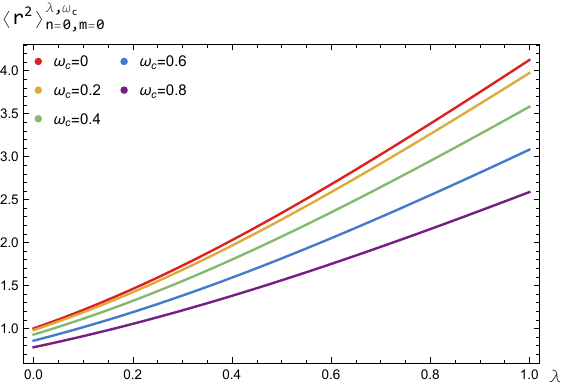}} &
\subfloat[]{\includegraphics[scale = 0.9]{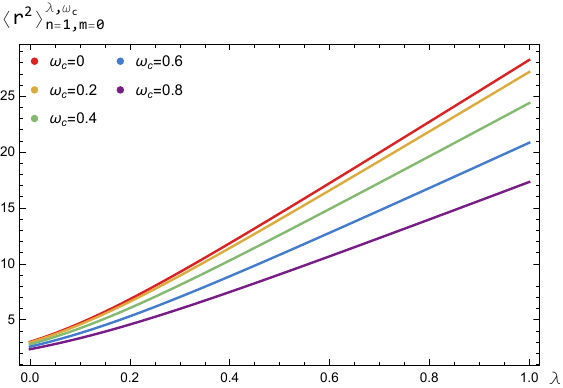}}
\end{tabular}
\caption{Expectation value of $r^2$ \eqref{eq: r2 darboux}  for the ground state (A) and the first excited state (B) as a function of $\lambda$, for different values of $\omega_c$. In all the cases $\omega=1$ and $l=0$.}
\label{grid:dispersion measures position}
\end{figure}

\begin{figure}[H]
\centering
\begin{tabular}{cccc}
\subfloat[]{\includegraphics[scale = 0.9]{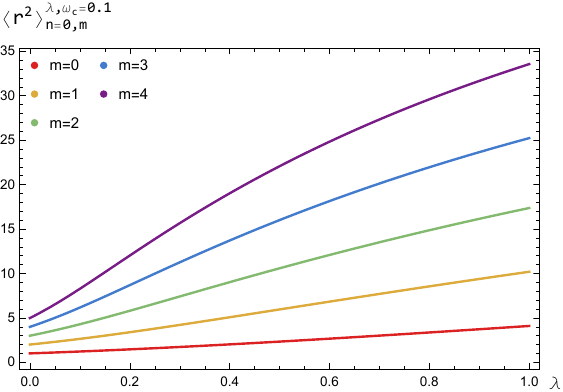}} &
\subfloat[]{\includegraphics[scale = 0.9]{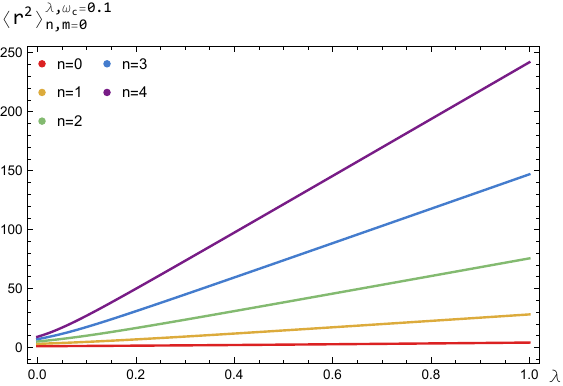}}
\end{tabular}
\caption{Expectation value of $r^2$ \eqref{eq: r2 darboux} for different values of $l$ and $n=0$ (A) or different values of $n$ and $l=0$ (B), as a function of $\lambda$. In all the cases $\omega=1$ and $\omega_c=0.1$.}
\label{grid:dispersion measures position diferentes cosas}
\end{figure}

As shown in Figure \ref{grid:dispersion measures position}, the expectation value of $r^2$ follows the same dependence as the localisation of the density in position space. This value increases with $\lambda$, and does so more rapidly as $n$ increases. However, the magnetic field tends to have the opposite effect. In terms of an oscillator, the oscillations become wider with $\lambda$, but are more confined by $\omega_c$. In Figure \ref{grid:dispersion measures position diferentes cosas}, we observe that both quantum numbers $n$ and $l$ also increase the expectation value. 

\subsubsection{Dispersion measures in momentum space}

The expected value of the momentum can be derived analytically by using the position representation of the momentum operator, namely
\begin{align}
\mean{p^2}_{n,m}^{\lambda,\omega_c} = - \int_0^{2\pi} \int_0^{\infty} \overline{\Psi_{n,m}^{\lambda,\omega_c}}(r, \varphi) \, \nabla^2 \Psi_{n,m}^{\lambda,\omega_c}(r, \varphi) \, r \, dr \, d\varphi
\end{align}
where the Laplacian in polar coordinates is given explicitly by
\begin{align}
\nabla^2 = \frac{1}{r} \frac{\partial}{\partial r} \left( r \frac{\partial}{\partial r} \right) + \frac{1}{r^2} \frac{\partial^2}{\partial \varphi^2} .
\end{align}
Integrating by parts we can simplify the expression to obtain
\begin{align}
\mean{p^2}_{n,m}^{\lambda,\omega_c} = \int_0^{2\pi} \int_0^{\infty} \abs{\nabla \Psi_{n,m}^{\lambda,\omega_c}(r, \varphi)}^2 \, r \, dr \, d\varphi=\int_0^{2\pi} \int_0^{\infty} \left[ 
\left| \frac{\partial \Psi_{n,m}^{\lambda,\omega_c}}{\partial r} \right|^2 
+ \frac{1}{r^2} \left| \frac{\partial \Psi_{n,m}^{\lambda,\omega_c}}{\partial \varphi} \right|^2 
\right] r \, dr \, d\varphi.
\end{align}
Since the wave function is radially symmetric we have that $\Psi_{n,m}^{\lambda,\omega_c}(r,\varphi)=R_{n,m}^{\lambda,\omega_c} \frac{1}{\sqrt{2\pi}} \expo{i m \varphi}$, where in addition $R_{n,m}^{\lambda,\omega_c}$ is a real function we have that
\begin{align} \notag
\mean{p^2}_{n,m}^{\lambda,\omega_c}  &=  \int_0^{2\pi} \int_0^{\infty} \left[ 
\left| \frac{\expo{i m \varphi}}{\sqrt{2\pi}}  \pdv{R_{n,m}^{\lambda,\omega_c}}{r} \right|^2 
+ \frac{1}{r^2} \left| R_{n,m}^{\lambda,\omega_c} \frac{1}{\sqrt{2\pi}} \pdv{\expo{i m \varphi}}{\varphi} \right|^2 
\right] r \, dr \, d\varphi, \\ \notag
&=\frac{1}{2\pi}\int_0^{2\pi} \int_0^{\infty} \left[ 
\paren{\pdv{R_{n,m}^{\lambda,\omega_c}}{r}}^2 
+ \frac{l^2}{r^2} \paren{R_{n,m}^{\lambda,\omega_c}}^2  
\right] r \, dr \, d\varphi, \\
&=\int_0^{\infty} \left[ 
\paren{\pdv{R_{n,m}^{\lambda,\omega_c}}{r}}^2 
+ \frac{l^2}{r^2} \paren{R_{n,m}^{\lambda,\omega_c}}^2  
\right] r \, dr.
\end{align}
We now separate these two integrals and define them as 
\begin{align}
    \mean{p^2_r}_{n,m}^{\lambda,\omega_c} &=\int_0^{\infty}  
 \paren{\pdv{R_{n,m}^{\lambda,\omega_c}}{r}}^2 
 r \, dr ,
    \\
    \mean{p^2_\varphi}_{n,m}^{\lambda,\omega_c} &=l^2 \int_0^{\infty} 
\frac{1}{r} \paren{R_{n,m}^{\lambda,\omega_c}}^2  \, dr,
\end{align}
where each term represents the expectation value of the radial and angular contributions of the Laplacian, respectively. Therefore,
\begin{equation}
    \mean{p^2}_{n,m}^{\lambda,\omega_c}=\mean{p^2_r}_{n,m}^{\lambda,\omega_c}+\mean{p^2_\varphi}_{n,m}^{\lambda,\omega_c}.
\end{equation}
The first integral can be divided into three terms, each with a different combination of Laguerre polynomials. By making $r \sqrt{\Otnl} =t$, we can write
\begin{equation}
    \mean{p^2_r}_{n,m}^{\lambda,\omega_c} =R_1+R_2+R_3 ,
\end{equation}
with
\begin{align}
    R_1 &=
    4 A
    \int_0^{\infty}  
    e^{-t^2}\, t^{3 + 2l}\, \paren{t^2 \lambda + \Otnl}\, \paren{L_{n - 1}^{1 + l} \paren{t^2}}^2
    \dt \\
    R_2&=
    A
    \int_0^\infty 
    \frac{
    e^{-t^2}\, t^{-1 + 2l}\, \paren{ -t^4 \lambda + t^2 \paren{\lambda + l \lambda - \Otnl} + l\, \Otnl }^2\, \paren{L_n^l \paren{t^2}}^2
    }{
    \paren{t^2 \lambda + \Otnl}
    } 
    \dt \\
    R_3&=
    4 A
    \int_0^\infty 
    e^{-t^2}\, t^{1 + 2l}\, \paren{t^4 \lambda - t^2 \paren{\lambda + l \lambda - \Otnl} - l\, \Otnl}\,
    L_{n - 1}^{1 + l} \paren{t^2}\, L_n^l \paren{t^2}
    \dt
\end{align}
where
\begin{equation}
    A= \frac{
    2\, \Otnl \, n!
    }{
    \paren{(1 + l + 2n)\lambda + \Otnl}\, \Gamma\paren{1 + l + n}
    } .
    \label{eq:constantRs}
\end{equation}
The first and third integrals integrals $R_1$ and $R_3$ can be computed using a combination of expected values for the Laguerre polynomials (Eq. \eqref{eq: laguerre expected}). Doing this, we obtain
\begin{align}
    R_1=\frac{4n\, \Otnl\, \paren{ \lambda(l + 2n) + \Otnl }}{ \lambda(l + 2n + 1) + \Otnl },
\end{align}
and
\begin{align}
    R_3 = -\frac{4n\, \Otnl\, \paren{ \lambda(l + 3n) + \Otnl }}{ \lambda(l + 2n + 1) + \Otnl }.
\end{align}
The integral $R_2$ is far more complex due to the $t$-dependence of the denominator. It can be computed as a sum of integrals of the form
\begin{align}
\int_0^\infty \frac{e^{-t^2} t^k}{\lambda t^2 + \Otnl} \,\mathrm{d}t 
= \frac{e^{\Otnl/\lambda} \Gamma\paren{ \frac{k+1}{2} } \paren{ \frac{\lambda}{\Otnl} }^{\frac{1}{2} - \frac{k}{2}} \Gamma\paren{ \frac{1}{2} - \frac{k}{2}, \frac{\Otnl}{\lambda} }}{2\lambda}.
\end{align}
In our case, since the closed form for the generalized Laguerre polynomials is
\begin{align}
    L_n^\alpha(x)=\sum_{i=0}^{n} (-1)^i \binom{n+\alpha}{n-i} \frac{x^i}{i!},
\end{align}
then the integral $R_2$ is 
\begin{equation}
    R_2=\sum_{i=0}^n \sum_{j=0}^n \frac{(-1)^i}{i!} \frac{(-1)^j}{j!} \binom{n+l}{n-i} \binom{n+l}{n-j} R_2^{(i,j)} ,
\end{equation}
where we have defined 
\begin{align}
    R_2^{(i,j)}&=
    A
    \int_0^\infty 
    \frac{
    e^{-t^2}\, t^{i+j-1 + 2l}\, \paren{ t^2 \paren{\lambda + l \lambda - \Otnl} + l\, \Otnl -t^4 \lambda}^2
    }{
    t^2 \lambda + \Otnl
    } 
    \dt ,
\end{align}
with $A$ given by Eq. \eqref{eq:constantRs}, and this integral being analytical. Summing up, the radial component can be written as 
\begin{align} \label{eq: mean p2 Darboux III magnetico}
\mean{p^2_r}_{n,m}^{\lambda,\omega_c}=\sum_{i=0}^n \sum_{j=0}^n \frac{(-1)^i}{i!} \frac{(-1)^j}{j!} \binom{n+l}{n-i} \binom{n+l}{n-j} R_2^{(i,j)}-\frac{4n^2\, \Otnl}{ \lambda(l + 2n + 1) + \Otnl } ,
\end{align}
where 
\begin{align} \notag
    R_2^{(i,j)}
    &=8\paren{\Otnl}^2 e^{\Otnl /\lambda} E_{\frac{k}{2}+l}\paren{\frac{\Otnl}{\lambda}} + \lambda\paren{ \lambda(k+2l)\paren{k^2+2k+4l+4} + 2\Otnl\paren{k^2+2k+4l-4} } \\
    &\times \frac{\Gamma(n+1)\Gamma\paren{\frac{k}{2}+l}}{8\Gamma(l+n+1)\paren{\lambda(l+2n+1)+\Otnl}},
\end{align}
with $k=i+j$ and where $E_{k+4}\paren{ \frac{\Otnl}{\lambda} }$ is the integral exponential function defined by
\begin{align}
    E_n(z) = \int_1^\infty \expo{-zt}/t^n \dt.
\end{align}
The integral for the angular contribution can be written again in terms of expected values of the Laguerre polynomials of same degree (see \eqref{eq: mean xs laguerre}), namely
\begin{align} 
    \mean{p^2_\varphi}_{n,m}^{\lambda,\omega_c} = \int_0^{\infty} 
\frac{l^2}{r^2} \paren{R_{n,m}^{\lambda,\omega_c}}^2 r \, dr &= l^2 \frac{\paren{\mathcal{N}^{\lambda,\omega_c}_{n,m}}^2}{\paren{\mathcal{N}_{n,m}^{\lambda,\omega_c}}^2} \paren{\mean{\frac{1}{r^2}}_{n,m}^{\lambda=0}+\mean{1}_{n,m}^{\lambda=0}} \\
&=\paren{\mathcal{N}^{\lambda,\omega_c}_{n,m}}^2  \frac{l^2 \lambda + l \Otnl}{2 \, n! \paren{\Otnl}^{1 + l}} \Gamma(n + l + 1),
\end{align}
which, by definition of the normalization constant in Eq. \eqref{eq: norm darboux magnetico}, is
\begin{align} \label{eq: mean p2 l magnetico}
     \mean{p^2_\varphi}_{n,m}^{\lambda,\omega_c}&=
      \frac{l \, \Otnl \, \paren{l \lambda + \Otnl}}{\paren{1 + l + 2n} \lambda + \Otnl} .
\end{align}
In the limit $\lambda \to 0$, these expressions simplify due to the limit of the product of the exponential integral and the exponential going to zero, recovering the Fock-Darwin results given by Eq. \eqref{eq: r2 p2 Fock-Darwin}. 

\begin{figure}[H]
\centering
\begin{tabular}{cccc}
\subfloat[]{\includegraphics[scale = 0.9]{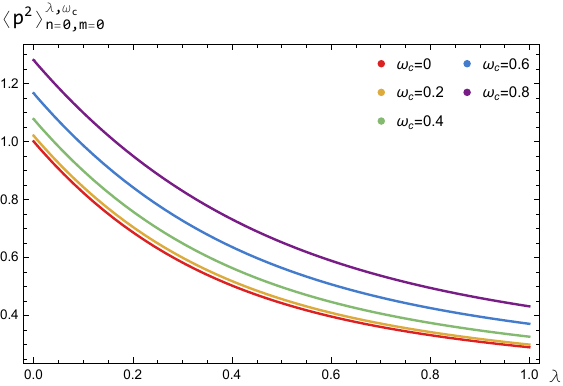}} &
\subfloat[]{\includegraphics[scale = 0.9]{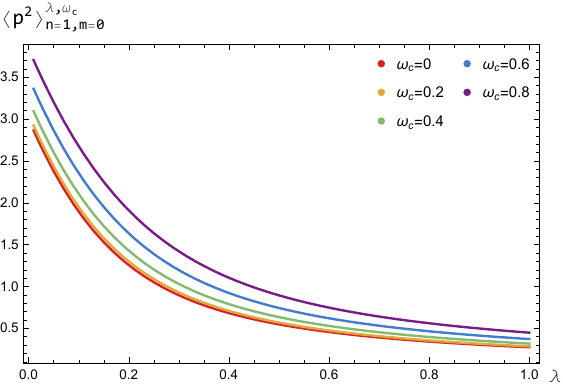}}
\end{tabular}
\caption{Expectation value of $p^2$ \eqref{eq: mean p2 Darboux III magnetico}+\eqref{eq: mean p2 l magnetico} for the ground state (A) and the first excited state (B) as functions of $\lambda$ ($\omega=1$ and $m=0$).}
\label{grid:dispersion measures momentum}
\end{figure}

\begin{figure}[H]
\centering
\begin{tabular}{cccc}
\subfloat[]{\includegraphics[scale = 0.9]{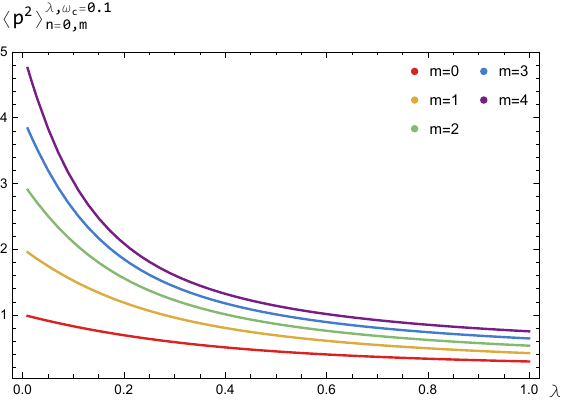}} &
\subfloat[]{\includegraphics[scale = 0.9]{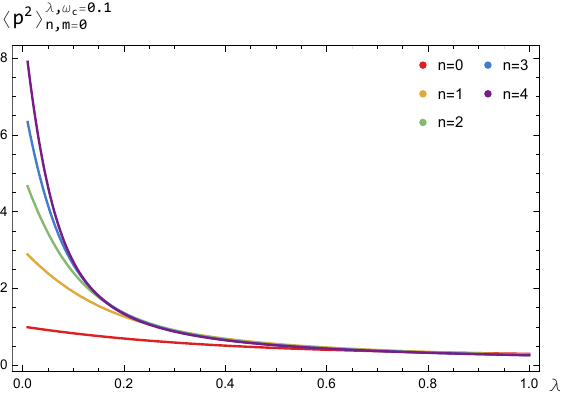}}
\end{tabular}
\caption{Expectation value of $p^2$ \eqref{eq: mean p2 Darboux III magnetico}+\eqref{eq: mean p2 l magnetico} for $n=0$ and different values of $m$ (A) or $m=0$ and different values of $n$ (B) as functions of $\lambda$ ($\omega=1$ and $\omega_c=0.1$).}
\label{grid:dispersion measures momentum distintas cosas}
\end{figure}

For $n=0$, the FDD results reduce to
\begin{align}
    \mean{p^2_r}_{0,m}^{\lambda,\omega_c}&=\Otol - \frac{ \paren{\Otol}^2 e^{\Otol/\lambda} E_{l+1} \paren{ \frac{\Otol}{\lambda} } }{ \lambda + \lambda l + \Otol }
\\
    \mean{p^2_\varphi}_{0,m}^{\lambda,\omega_c}&=\frac{l \, \Otol \, \paren{l \lambda + \Otol}}{\paren{1 + l} \lambda + \Otol}, \\
    \mean{p^2}_{n,m}^{\lambda,\omega_c}&=\frac{ \Otol\, \paren{ -\Otol e^{\Otol/\lambda} E_{l+1}\paren{ \frac{\Otol}{\lambda} } + \lambda \paren{ l^2 + l + 1 } + l\, \Otol + \Otol } }{ \lambda + \lambda l + \Otol } .
\end{align}
And for the ground state, $n=l=0$, 
\begin{align} \label{eq: p2 ground state}
\mean{p^2_\varphi}_{0,0}^{\lambda,\omega_c}&=0, \\
\mean{p^2}_{0,0}^{\lambda,\omega_c}&=\mean{p^2_r}_{0,0}^{\lambda=0}=\Otoo - \frac{ \paren{\Otoo}^2 e^{\Otoo/\lambda} E_1\paren{ \frac{\Otoo}{\lambda} } }{ \lambda + \Otoo },
\end{align}
it is easily noticeable (see Figure \ref{grid:dispersion measures momentum}) that as the curvature parameter $\lambda$ increases, the expected value of the momentum decreases, as one would expect from the computation of the entropies. In Figure \ref{grid:dispersion measures momentum distintas cosas}, we observe the same trend as in Figure \ref{grid:dispersion measures position diferentes cosas}: the expectation value increases with both quantum numbers $n$ and $l$.

\subsubsection{Dispersion-based uncertainty principle}

\begin{figure}[H]
\centering
\begin{tabular}{cccc}
\subfloat[]
{\includegraphics[scale = 0.9]{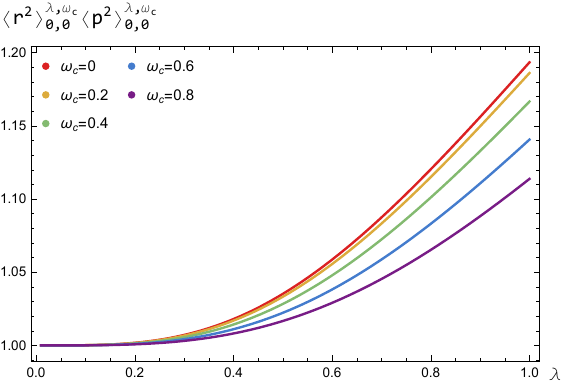}} &
\subfloat[]{\includegraphics[scale = 0.9]{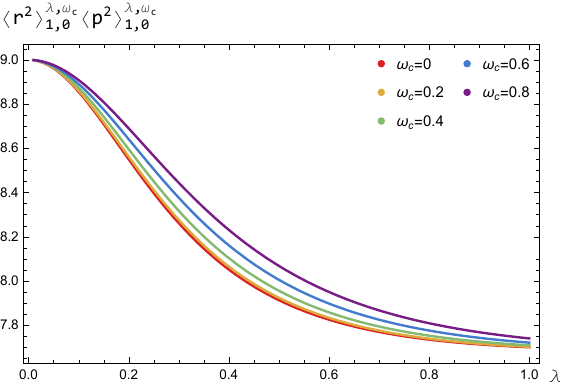}}
\end{tabular}
\caption{Uncertainty principle for the ground state (A) and the first excited state (B), as a function  of $\lambda$ ($\omega=1$ and $l=0$).}
\label{grid:dispersion measures}
\end{figure}

In order to see the effects of the interplay between the curvature parameter and the magnetic field, in Figure \ref{grid:dispersion measures} we have plotted the product given in \eqref{eq:dispersion_UP} as a function of $\lambda$, for different values of $\omega_c$. From these plots, the opposing influence of $\lambda$ and $\omega_c$ is apparent once again.

In Figure \ref{grid:dispersion measures} (A), we can see how the uncertainty increases with $\lambda$ and decreases with $\omega_c$. In the limit $\lambda \to 0$, the uncertainty principle for the harmonic oscillator is recovered, and thus the product $\mean{x}_{0,0}^{\lambda,\omega_c}\mean{p}_{0,0}^{\lambda,\omega_c}$ saturates to $1$, as shown in Eq. \eqref{eq:dispersion_UP0}, regardless of the value of $\omega_c$. 

Moreover, while the uncertainty increases with $\lambda$ for the ground state, it decreases for the excited states as shown in Figure \ref{grid:dispersion measures} (B) for $n=1$. In any case, it is clear that the value of $\omega_c$ is opposed to the effect of $\lambda$ for $\lambda \neq 0$. 

\begin{figure}[H]
\centering
\begin{tabular}{cccc}
\subfloat[]{\includegraphics[scale = 0.9]{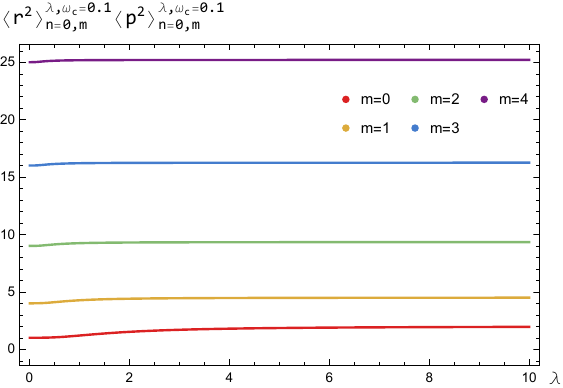}} &
\subfloat[]{\includegraphics[scale = 0.9]{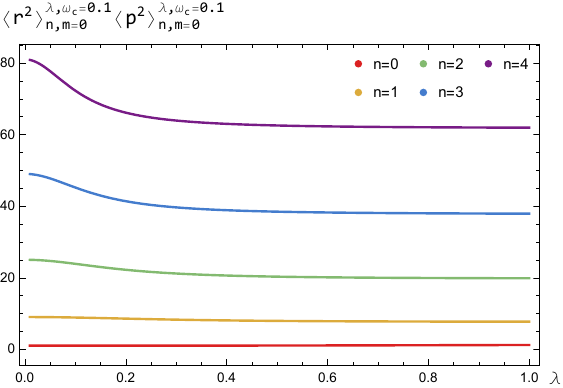}}
\end{tabular}
\caption{Uncertainty principle for $n=0$ and different values of $l$ (A) or $l=0$ and different values of $n$ (B) as a function of $\lambda$. In all cases $\omega=1$ and $\omega_c=0.1$.}
\label{grid:dispersion incertidumbre distintas cosas}
\end{figure}

For $n=0$, the uncertainty principle saturates in the limit $\lambda \to 0$ to $\paren{l+1}^2$, as shown in Eq. \eqref{eq:dispersion_UP}. This behaviour can be observed in Figure \ref{grid:dispersion incertidumbre distintas cosas} (A), where it is shown that for $l=0,1,2,3,4$, the uncertainty principle saturates to $1,4,9,16,25$, respectively. Since this figure is plotted for $n=0$, the uncertainty slightly increases with $\lambda$ in all cases. However, Figure \ref{grid:dispersion incertidumbre distintas cosas} (B) shows that the uncertainty decreases for the remaining states. This decrease becomes more pronounced as $n$ becomes larger. Note that, although the scaling makes this harder to see for smaller values of $n$, it can be appreciated in combination with Figure \ref{grid:dispersion measures}. The red line $(n=0)$ in Figure \ref{grid:dispersion incertidumbre distintas cosas}(B) could be added to panel A of Figure \ref{grid:dispersion measures}, while the yellow line $(n=1)$ could be added to panel B of the same figure.

\section{Magnetic field regimes of special interest} \label{sec: Magnetic field regimes of special interest}

In this section, we study in more detail the interplay between the curvature parameter and the magnetic field. 

\subsection{Counteracting curvature effects via magnetic field coupling} \label{sec: Counteracting curvature effects via magnetic field coupling}

\begin{figure}[H]
\begin{tabular}{cccc}
\subfloat[]{\includegraphics[scale = 0.9]{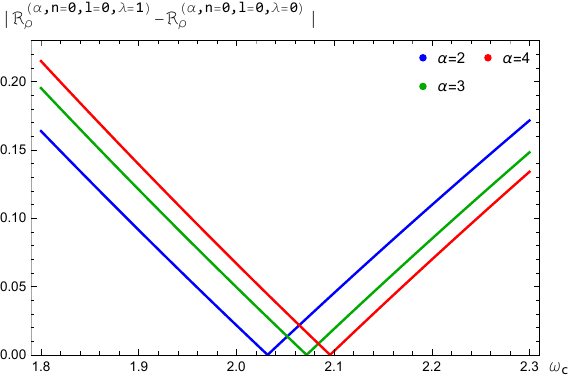}} &
\subfloat[]{\includegraphics[scale = 0.9]{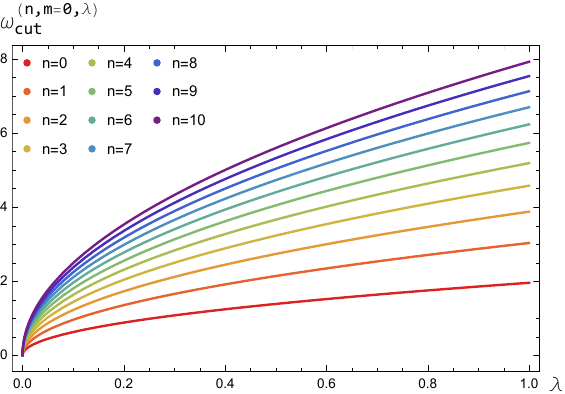}} 
\end{tabular}
\caption{(A) Difference of the Rényi entropies of the FDD with $\lambda=1$ and of the harmonic oscillator as a function of $\omega_c$ (with $\omega=1$, $n=l=0$ and parameters $\pe=2,3,4$). The value of $\omega_c$ that makes entropies equal (intersection with the x-axis) changes with $\pe$. (B) $\omega^{(n,l,\lambda)}_{\mathrm{cut}}$ \eqref{eq: omega cut en r} as a function of $\lambda$ (with $\omega=1$ and $l=0$).}
\label{grid: omegacutr}
\end{figure}

In the previous Section, we showed that $\lambda$ and $\omega_c$ have opposite effects, and thus it is natural to ask whether an appropriate balance between them can restore the behaviour of the harmonic oscillator. Since both the Rényi and Tsallis entropies depend on the parameter $\alpha$ (and reduce to the Shannon entropy in the limit $\alpha \to 1$), it is not possible to find a universal pair of values of $\lambda$ and $\omega_c$ that recovers all the information entropies of the harmonic oscillator. This is shown in Figure~\ref{grid: omegacutr}~(A), where the entropy difference reaches zero at three different combinations of $\omega_c$ and $\lambda$, each corresponding to a different value of $\pe$.

However, for the position expectation value, there is indeed a value of the Larmor frequency $\omega_c$ for which the net contribution of the curvature and magnetic fields vanishes. This value can be obtained by equating the two expressions for the position expectation value given in Eqs.~\eqref{eq: r2 p2 Fock-Darwin} and~\eqref{eq: r2 darboux}.
\begin{align}
   \frac{1}{\Ot+(2n+l+1) \lambda} \paren{2n+l+1 +\frac{\lambda}{\Ot} \paren{l^2+l (6 n+3)+6 n (n+1)+2}}=
    \frac{l+2n+1}{\omega}.
\end{align}
We denote this value of the Larmor frequency as $\omega^{(n,m,\lambda)}_{\mathrm{cut}}$, and it reads
\begin{align} \label{eq: omega cut en r}
     \omega^{(n,m,\lambda)}_{\mathrm{cut}}=\sqrt{\lambda^2 m^2+\paren{2n+m+1}\lambda g_{nm}^\lambda+\frac{\paren{g_{nm}^\lambda}^2}{4}-\omega^2}-\lambda m,
\end{align}
where we have defined
\begin{align}
c_{nl}&=2n+l+1, \\
    g_{nm}^{\lambda}&=\sqrt{c_{nl}^2\lambda^2+\frac{2\lambda\omega\paren{l\paren{l+8n+4}+8n\paren{n+1}+3}}{c_{nl}}+\omega^2}-c_{nl}\lambda+\omega.
\end{align}
For $\lambda=0$, expression~\eqref{eq: omega cut en r} goes to zero, as shown in Figure~\ref{grid: omegacutr}~(B). This is because the trivial case $\lambda=\omega_c=0$ corresponds to the harmonic oscillator limit. Moreover, we show that the value of $\omega_c$ required to cancel the effect of the curvature parameter $\lambda$ increases not only with $\lambda$ but also with $n$. This is due to the interplay between $n$ and $\lambda$: the larger $n$, the stronger the effect of $\lambda$. It is important to remark that the fact that the effect of $\lambda$ can be counteracted by $\omega_c$ in position space does not imply that the same cancellation occurs simultaneously in momentum space.

\begin{figure}[H]
    \centering
    \includegraphics[width=0.5\linewidth]{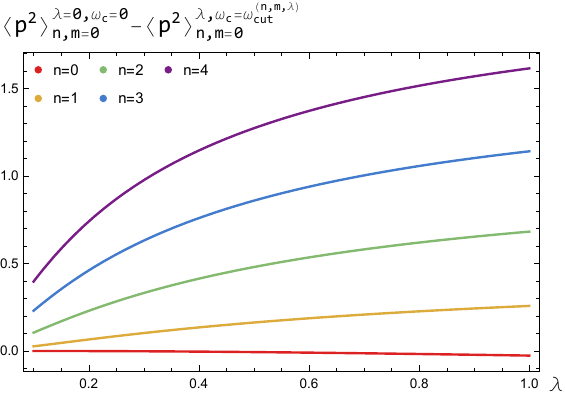}
    \caption{Difference between the expectation values of $p^2$ for the FDD and the harmonic oscillator, with the same expectation values of $r^2$ in both systems for different values of $n$ ($\omega=1$, $m=0$).}
    \label{fig:p2 con esos valores}
\end{figure}

To show this, we use the values of $\omega_{\mathrm{cut}}^{(n,l,\lambda)}$ obtained from the cancellation condition in position space and evaluate the corresponding expectation value of $p^2$. If the compensation were also valid in momentum space, the difference between this value and the harmonic oscillator expectation value would vanish. However, as shown in Figure~\ref{fig:p2 con esos valores}, this difference is not zero. Therefore, the balance between curvature and magnetic field that restores the harmonic oscillator behaviour in position space does not restore it in momentum space. 

In the same figure, it is also apparent that the difference becomes larger as $n$ increases. This suggests that the interplay between $\lambda$ and $n$ is different in position and momentum spaces. This difference helps explain why the effects of $\lambda$ cannot be counteracted simultaneously in both spaces by tuning $\omega_c$. In conclusion, we have shown that the opposite trends induced by curvature and by the magnetic field do not imply a complete mutual cancellation. Even when $\lambda$ and $\omega_c$ are tuned to balance a given observable, the system does not recover the full behaviour of the undeformed harmonic oscillator.

\subsection{Inverted magnetic field direction}

Another interesting question is what happens when inverting the direction of the magnetic field. The effect of this is that the Larmor frequency changes from $\omega_c$ to $-\omega_c$ in the expression of the FDD system frequency $\Ot$. Let us remind that this expression reads
\begin{align} \notag
   \Ot&= 
   \sqrt{-2 \lambda  \hbar  \left(\sqrt{c_{n,l}^2 \left(\omega _c \left(\omega _c+2 \lambda  l \hbar \right)+\lambda ^2 \hbar ^2 c_{n,l}^2+\omega ^2\right)}-l \omega _c-\lambda  \hbar  c_{n,l}^2\right)+\omega _c^2+\omega ^2},
\end{align}
where we define $c_{n,l}=2n+l+1$ for simplicity. For states with $l=0$, the FDD frequency reduces to
\begin{align}
    \Omega_{n,0}^{\lambda ,\omega_c}=\sqrt{2 \lambda  \hbar  \left(\lambda  (2 n+1)^2 \hbar -\sqrt{(2 n+1)^2 \left(\omega _c^2+\lambda ^2 (2 n+1)^2 \hbar ^2+\omega ^2\right)}\right)+\omega _c^2+\omega ^2} .
\end{align}
This is an even function of $\omega_c$, namely
\begin{align} \label{eq: omega cut neg l=0}
    \Omega_{n,0}^{\lambda,\omega_c}(-\omega_c)=  \Omega_{n,0}^{\lambda,\omega_c}(\omega_c) ,
\end{align}
which implies that the system remains unchanged for positive and negative values of $\omega_c$ if $l=0$. Therefore, all previous results remain valid under an inversion of the magnetic field.

\begin{figure}[H]
\begin{tabular}{cccc}
\subfloat[]{\includegraphics[scale = 0.9]{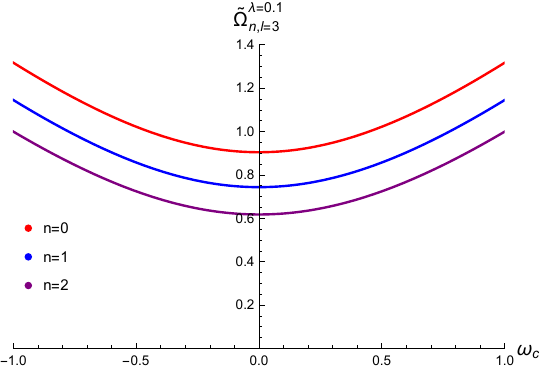}} &
\subfloat[]{\includegraphics[scale = 0.9]{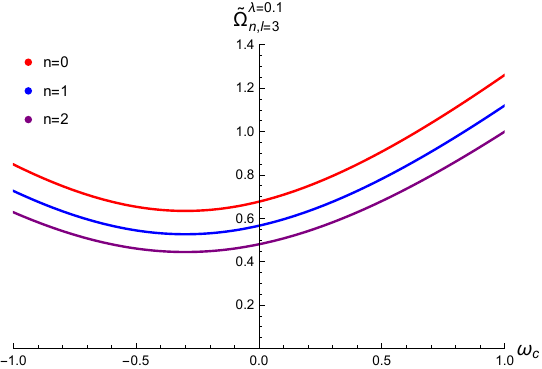}} 
\end{tabular}
\caption{Symmetric $(l=0)$ and asymmetric $(l=3\neq0)$ values of $\Ot$ for $n=0,1,2$, $\lambda=0.1$ and $\omega=1$.}
\label{grid: omegacutr sym}
\end{figure}

However, when $l \neq 0$, an asymmetry arises between the positive and `negative' values of $\omega_c$. For the latter, we can solve the equation
\begin{align}
    \Ot \paren{-\omega_{c}}=\Ot \paren{\omega_{c}+f(\lambda,l)} ,
\end{align}
where $f(\lambda,l)$ is the function we need to add to make the value of $\omega_c$  positive and still obtain the same frequency $\Ot$. This equation admits a very simple solution, namely
\begin{align}
    f(\lambda,l)=-2 \lambda l \hbar.
\end{align}
In conclusion, when inverting the magnetic field, it is as if the value of $\omega_c$ were negative. But this is equivalent in the wave function to a positive value of the frequency adding a corrective term due to the magnetic field inversion,
\begin{align} \label{eq: relacion invertir campo}
    -\omega_{c} \rightarrow \omega_{c}-2 \lambda  l \hbar .
\end{align}
As expected, for $l = 0$, the term $-2 \lambda l \hbar$ vanishes, and one recovers Eq.~(\ref{eq: omega cut neg l=0}).  Therefore, all the entropy plots presented in Section \ref{sec: resultados FDD}, which were obtained for $l=0$, remain valid when the direction of the magnetic field is inverted.

\section{Concluding remarks}  \label{Conclusions magnetico}

Exactly solvable nonlinear quantum models are scarce but relevant, since their analytical solutions allows a deep investigation of all their mathematical and physical properties, in particular their information theoretic features. In this paper we have analysed in detail the Fock-Darwin-Darboux (FDD) system, a novel exactly solvable two-dimensional nonlinear quantum system defined by a charged particle moving on the Darboux III surface (with variable negative curvature) and subjected  to both a magnetic field and an isotropic oscillator potential. This system is the generalisation of the renowned Fock-Darwin (FD) system, which is the flat limit of the FDD Hamiltonian.

The FDD system has three essential parameters: the curvature parameter $\lambda$, the magnetic field strength $B$ (or its associated Larmor frequency $\omega_c$) and the oscillator strength $k$ (or the corresponding oscillator frequency $\omega$). The exact solvability of the FDD system makes it possible to study the non trivial interplay among these three parameters in all contexts. In particular, we have focused in the analysis of the Shanon, Rényi and Tsallis entropies of the FDD system, as well as of its dispersion measures. When possible, analytical expressions have been found, and the flat $\lambda\to 0$ limit of all these quantities leads to the results corresponding to the FD system, which we have also presented for the first time (Section~\ref{sec: Fock-Darwin quantum system}). In both systems, the magnetic field increases the effective frequency $\Ot$, whereas the parameter $\lambda$ decreases the effective frequency, thereby counteracting the influence of the field.

In Section~\ref{sec: resultados FDD} we computed both entropic and dispersion measures. Although we presented the Shannon entropy for this system, it just corresponds to the $N=2$ case with a modified frequency of the general result in \cite{ballesteros2023shannon}. For this reason, particular emphasis was placed on the Rényi and Tsallis entropies, for which we derived closed-form relations (Eqs.~\ref{eq: renyi magnetic darboux}) and (\ref{eq: tsallis magnetic darboux}) in position space, expressed in terms of the quantum numbers and system parameters. In addition, a numerical study was performed for the momentum-space entropy and the entropy-based uncertainty principle. The mean values of the radial coordinate were obtained analytically on both spaces (see Eqs.~\ref{eq: rk darboux}), (\ref{eq: r2 darboux}), (\ref{eq: mean p2 Darboux III magnetico}), (\ref{eq: mean p2 l magnetico})). Overall, both entropic and dispersion measures were found to increase with $\lambda$ in position space and to decrease in momentum space, while the magnetic-field parameter $\omega_c$ produced the opposite effect, and the same can be observed in all uncertainty relations. Consistently, in the limit $\lambda \to 0$, the uncertainty relations no longer depend on the magnetic field, as expected from the Fock-Darwin system.

In Section~\ref{sec: Magnetic Darboux III oscillator} we also showed that, despite their competing effects, the harmonic oscillator cannot be fully recovered by jointly tuning $\lambda$ and $\omega_c$. The entropic measures cannot be matched in either space, while the dispersion measures can coincide in one space or the other, but not simultaneously. In other words, it is possible to confine a particle to the same expected position value, but its dynamics will be slower (see Figure~\ref{fig:p2 con esos valores}). In the same section, we also considered inversion of the magnetic-field direction. This modification changes the value of $\Ot$ proportionally to the angular momentum quantum number $l$ and the curvature parameter $\lambda$. For $\lambda \to 0$, however, field inversion has no impact on either the entropic or dispersion measures.

Finally, we investigated regimes in which one parameter dominates over the other. Owing to the structure of the term $1+\lambda r^2$, the probability density can be separated into two contributions, each becoming negligible when the opposite parameter is large. For strong curvature, the Darboux III dynamics prevail, allowing for an approximation of the Fourier transform that improves with increasing excitation of the state. This approximation could be particularly useful in the study of highly excited, Rydberg-like states. Conversely, when the magnetic field dominates, the dynamics reduce to those of the Fock–Darwin oscillator.

It is worth emphasizing that the Darboux III oscillator combines the advantage of closed-form expressions for all energy levels and wave functions with the presence of a nonlinearity parameter $\lambda$. In this work, we have carried out a combined analytical and numerical study of the magnetic field in this system, thereby enriching the understanding of quantum-information measures in nonlinear quantum systems under external fields. Beyond the two-dimensional case, our approach naturally extends to higher-dimensional Darboux III oscillators, providing a fertile framework for generalizing the entropic and dispersion measures obtained here. Finally, it would be of particular interest to examine the model with negative $\lambda$ parameter, corresponding to different curvature properties. These lines of research are currently in progress and will be reported elsewhere.

\section*{Acknowledgements}

The authors acknowledge partial support from the grant PID2023-148373NB-I00 funded by MCIN/AEI /  10.13039/501100011033 / FEDER -- UE, and the Q-CAYLE Project funded by the Regional Government of Castilla y León (Junta de Castilla y León) and the Ministry of Science and Innovation MICIN through NextGenerationEU (PRTR C17.I1).

\end{document}